\documentclass[showpacs,twocolumn,prd,amsmath,nofootinbib,amssymb]{revtex4}
\usepackage{graphicx}
\usepackage{subfigure}
\usepackage{hyperref}
\usepackage{amssymb}
\usepackage[latin1]{inputenc}
\usepackage[english]{babel}
\usepackage{epsfig}
\usepackage{wasysym}
\usepackage{color,xcolor}
\usepackage{pdflscape}
\usepackage{amsmath}
\usepackage{bm}
\usepackage{epsfig}
\usepackage{amsfonts}
\usepackage{dcolumn}

\definecolor{purple}{rgb}{1,0,1}
\definecolor{lime}{HTML}{A6CE39} 

\begin{document}

\title{Regular multihorizon black holes in General Relativity\\}

\author{ Manuel E. Rodrigues$^{(a,b)}$\footnote{E-mail address: 
esialg@gmail.com}, Marcos V. de S. Silva$^{(b)}$\footnote{E-mail 
address: marco2s303@gmail.com}, Andrew S. de Siqueira$^{(b)}$\footnote{ansiquei@yahoo.com.br}}
\affiliation{$^{(a)}$Faculdade de Ci\^{e}ncias Exatas e Tecnologia, 
Universidade Federal do Par\'{a}\\
Campus Universit\'{a}rio de Abaetetuba, 68440-000, Abaetetuba, Par\'{a}, 
Brazil\\
$^{(b)}$Faculdade de F\'{\i}sica, Programa de P\'os-Gradua\c{c}\~ao em 
F\'isica, Universidade Federal do 
 Par\'{a}, 66075-110, Bel\'{e}m, Par\'{a}, Brazil}


\begin{abstract}
In this work, new solutions for regular black holes that have multihorizons are proposed. These are formed by the direct product of solutions already published in the literature, which are described through the coupling of gravity with nonlinear electrodynamics. We analyze the regularity of the spacetime, the electric field, and the energy conditions of each solution. The strong energy condition is always violated within the event horizon in all solutions, while other energy conditions depend on the ratio between extreme charges of isolated solutions. For solutions with four horizons, we present two examples, Bardeen-Culetu and Balart-Culetu. Both solutions are regular, but the first do not satisfy all the energy conditions, except the strong, because it has an extreme charge ratio of 1.57581, great value. The second solution, on the other hand, can satisfy all other energy conditions, except the SEC, and has an extreme charge ratio of 1.09915, a value that allows this feature. It's also proposed a regular solution with up to six horizons, Balart-Culetu-Dymnikova, where, for a given charge value, we can verify that it satisfies all energy conditions, except the strong one. This was possible due to the ratio between extreme charges that are neither too high nor too close. We propose solutions with any number of horizons. We show that points where $-F(r)$ has a non null minimum represent a cusp in the Lagrangian $-L(F)$. We also show an example of multihorizon solution with magnetic charge. Multihorizon solutions may exhibit exotic properties, such as negative energy density, or violation of energy conditions, but which can be circumvented with a selected choice of customized solutions and extreme charge values, resulting in regular black hole solutions that satisfy all energy conditions, less the strong.
\end{abstract}

\date{\today}

\maketitle



\section{Introduction}
\label{sec1}
Classical mechanics, despite successfully describing many phenomena, requires the existence of an inertial framework which cannot be precisely defined. Another problem that arises in the classical context is the fact that Maxwell equations are not invariant by Galileo transformations \cite{Inverno}. Being related to classical mechanics, Newton gravity also had difficulties in describing some phenomena. This motivated Albert Einstein to develop the theory of general relativity, published in 1915 \cite{RobertWald}, in which gravity is no longer a force, but the spacetime geometry itself, this being a pseudo-Riemannian manifold, whose equations of motion are coupled nonlinear differential equations. Einstein did not solve these equations, it was Karl Schwarzschild, in 1916, who obtained the first solution, a vacuum solution with spherical symmetry \cite{Schwarzschild}. This solution has a null hyper-surface, called event horizon and a point where spacetime is singular. Over the years after Schwarzschild solution to the present day, many other solutions have been found, between the various solutions, we can highlight two, the Reissner-Nordstrom solution and the one proposed by J. M. Bardeen \cite{BardeenConf}. Both solutions have two horizons, but the first has mass and electric charge, that can be obtained from the coupling of general relativity and Maxwell electromagnetic theory, and presents a singularity; the second can be built to have mass and electric charge from the coupling of general relativity with nonlinear electrodynamics, but it has no singularity at any point in spacetime. The Bardeen solution motivates the search for new solutions that do not present singularities in the causal structure, regular solutions. A convenient way to obtain regular black hole solutions without rotation is to use general relativity coupled to nonlinear electrodynamics with spherical symmetry, in which case the stress-energy tensor is diagonal and has symmetry $T^2_{\ 2} = T^3_{\ 3}$ and $T^0_{\ 0} = T^1_{\ 1}$, this implies that there will be only two equations of motion that are linearly independent, and the matter, at the center of the radial coordinate, will have a de Sitter-like behavior,
pressure = -density.

Maxwell linear electrodynamics is characterized by the following properties: it is invariant by the gauge transformation of group $U(1)$ and has second-order linear equations in the potentials. These are excellent properties, but when relaxed, new phenomena arise that do not appear in the linear case. For example, if we break $U(1)$ symmetry invariance, equations of an order greater than two in the potentials appear, as in the case of Podolsky and Proca electrodynamics \cite{Podolsky:1942zz,Bopp,Proca:1900nv}. If we relax for nonlinear equations, several families of electrodynamics appear with many other new phenomena, such as linear magnetic birefringence, where the propagation of light is anisotropic, depending on which direction on the polarization of light and the external magnetic field is applied \cite{Cadene}. These classes of electrodynamics are commonly called nonlinear electrodynamics (NED).

The first description of NED, in 1934, was formulated by Born and Infeld \cite{BORN}, when they wanted to eliminate the effects of the self-energy on the fields of a charged particle, and the singularity that appeared in the description at the point above the charge. This formulation corrects these difficulties. Two years later, in 1936, Euler and Heisenberg investigated photon-photon scattering and formulated another description of NED \cite{HEISENBERG}. Three years after Born and Infeld (BI) proposed their electrodynamics, Hoffmann coupled the Lagrangian of BI to gravitation \cite{HOFFMANN}. Over the years, several applications have appeared for the NEDs, here is a list of some of them: ionization of the hydrogen atom \cite{hidrogenio}, baryogenesis \cite{bariogenese}, CMB polarization \cite{polariza}, multicooling \cite{multirrefrigencia}, neutrino astrophysics \cite{neutrinos}, light propagation in one direction \cite{propagacao}, pulsar \cite{pulsar}, cosmological inflation \cite{inflacao}, photon gas thermodynamics \cite{termodinamica}, acceleration of the Universe \cite{aceleracao}. NED was measured in the laboratory by the following experiments: PVLAS (Polarizzazione del Vuoto by LASERs) \cite{PVLAS}, LSW (Light Shining through Walls) \cite{LSW}, BMV (Biréfringence Magnétique du Vide / Toulouse) \cite{Bailly}, VH (Photon Collider = Vacuum Hohlraum) \cite{Pike}, XFELS (X-ray Free Electron LASERS) \cite{XFELS}, ELI (Extreme Light Infrastructure) \cite{ELI}, SULF (Shanghai Ultra-Laser Facilities) \cite{SULF}; and may have proof corroborated by the experiments: SEL (Station of Extreme Light, Shanghai, 2023) XCELS (ExaWatt Center for Extreme Light Studies, Russia, 2026).

With the proposal to regularize the electric field provided by a point charge, at the origin of the radial coordinate, and Bardeen's proposal for a regular solution in all space-time, some authors considered to use NED to describe regular solutions. Pellicer and Torrence \cite{Pelicer} used Plebanski's NED to obtain a spherically symmetrical and regular solution. Beato and Garcia also used the NED to formulate the Bardeen solution as a nonlinear magnetic monopole \cite{Beato-Garcia}. This particular solution is not asymptotically Reissner-Nordstrom. So, we have some other solutions regulars, such as de Bronnikov \cite{Bronnikov}, Dymnikova \cite{Dymnikova}, Culetu \cite{Culetu}, and Balart and Vagenas \cite{Balart}.

The work of Odintsov and Nojiri's \cite{Odintsov-Nojiri} deals with new solutions of regular de Sitter-type black holes with multihorizons in General Relativity, $f(R)$ gravity, and Gauss-Bonnet in $5D$. In the same way, the article by Gao et al \cite{Gao} brings new solutions for black holes with multihorizons. Thus, Rodrigues and Silva formulate regular black holes with multihorizons in modified gravity theory $f(G)$ \cite{gravidade_modificada}. So, the natural question is ``can we formulate new solutions for regular black holes with multihorizons in General Relativity''. The main objective of this work is to answer this question and analyze these possible new solutions.

The structure of this paper is organized as follows. In Sec. \eqref{sec2}, we present regular solution in general relativity and some features about these solutions as energy conditions, regularity and electric field. In Sec. \eqref{sec3}, we present the multihorizon black hole solution and a method to build regular black holes with multiple horizons. Our conclusions and perspectives are in Sec. \eqref{sec5}. We adopt $c=G=1$.

\section{Regular black hole in General Relativity}
\label{sec2}

Regular black holes can be interpreted as solutions of Einstein equations with nonlinear electrodynamics. The action that describes this type of theory is given by
\begin{equation}
S=\int d^4x\left[R+\kappa^2 L(F)\right],
\label{EHA}
\end{equation}
where $R$ is the curvature scalar, $\kappa^2=8\pi$, and $F=\frac{1}{4}F^{\mu\nu}F_{\mu\nu}$ is the electromagnetic scalar, with $F_{\mu\nu}=\partial_\mu A_\nu-\partial_\nu A_\mu$ being the Maxwell-Faraday tensor. If we vary the action \eqref{EHA} with respect to $g_{\mu\nu}$ and $A_\mu$, we get
\begin{eqnarray}
R_{\mu\nu}-\frac{1}{2}g_{\mu\nu}R=\kappa^2T_{\mu\nu},\label{EE}\\
\nabla_{\mu}\left[L_FF^{\mu\nu}\right]=\partial_\mu\left[\sqrt{-g}L_FF^{\mu\nu}\right]=0,\label{ME}
\end{eqnarray}
where $R_{\mu\nu}$ is the Ricci tensor and $T_{\mu\nu}$ is the stress-energy tensor, given by
\begin{equation}
T_{\mu\nu}=\frac{1}{8\pi}\left[g_{\mu\nu}L(F)-L_F F_{\mu}^{\ \alpha}F_{\nu\alpha}\right],
\end{equation}
with $L_F=\partial L(F)/\partial F$.

Let us consider a spherically symmetric and static spacetime described by the line element
\begin{equation}
ds^2=f(r)dt^2-f(r)^{-1}dr^2-r^2\left(d\theta^2+\sin^2\theta d\phi^2\right).\label{le}
\end{equation}
If the source has only electric charge, we may integrate the modified Maxwell equation \eqref{ME}, to the line element \eqref{le}, and we find that the only nonzero and independent component of $F^{\mu\nu}$ is
\begin{equation}
F^{10}=\frac{q}{r^2}L_F^{-1}.\label{ef}
\end{equation}

Using the line element \eqref{le} and the electric field \eqref{ef}, the nonzero components of the Einstein equations are
\begin{eqnarray}
	\frac{1}{r^2}-\frac{f(r)}{r^2}-\frac{f'(r)}{r}&=&\left[L(F)+\frac{q^2}{r^4}L_F^{-1}\right],\label{eq1}\\
	-\frac{f'(r)}{r}-\frac{f''(r)}{2}&=&L(F).\label{eq2}
\end{eqnarray}

To regular solutions, we may write $f(r)$ as
\begin{equation}
f(r)=1-\frac{2M(r)}{r},
\end{equation}
where the mass function, $M=M(r)$, must satisfy the conditions $\lim\limits_{r\rightarrow 0}M/r=0$, to guarantee the regularity, and $\lim\limits_{r\rightarrow \infty}M=m$, where $m$ is de Arnowitt-Deser-Misner (ADM) mass. Using the equations of motion \eqref{eq1} and \eqref{eq2}, we find
\begin{eqnarray}
L(r)&=&-\frac{2f'+rf''}{2r},\label{lel}\\
L_F(r)&=&\frac{2q^2}{r^2\left(2-2f(r)+r^2f''(r)\right)}\label{lf}.
\end{eqnarray}
Therefore, for which regular black hole model, we will have a different nonlinear electrodynamics. The electromagnetic scalar $F$ is
\begin{equation}
F(r)=-\frac{1}{2}\left[F^{10}(r)\right]^2=-\frac{\left(2-2f+r^2f''\right)^2}{8q^2}\label{F}.
\end{equation}
In \cite{Bronnikov,Bronnikov:2000yz,Bronnikov:2019fgh}, Bronnikov presented a theorem of non-existence which says that the electrodynamic of a static regular solution with electric charge may not behave asymptotically as Maxwell to $F\rightarrow 0$ at $r\rightarrow 0$. If the electrodynamic behaves as Maxwell to weak fields, we have $L(F)\rightarrow F$ and $L_F\rightarrow 1$ to $F\rightarrow 0$. Usually, to regular solutions with electric charge, $F\rightarrow 0$ at $r\rightarrow 0$ and $r\rightarrow \infty$. To guarantee the regularity in the center, we need $FL_F<\infty$ while $FL_F^2\rightarrow -\infty$, which implies in $L_F\rightarrow \infty$ with $F\rightarrow 0$ at $r\rightarrow 0$, so that, the solution may not have a regular center with an electrodynamics which behaves like Maxwell in this region. However, the electromagnetic theory may behaves as Maxwell at $r\rightarrow \infty$. So that, to the same value of $F$, we may have different $L(F)$. 

We may also define the dual tensor $P_{\mu\nu}=L_FF_{\mu\nu}$. Using \eqref{ef} and \eqref{F}, we have the following scalar 
\begin{eqnarray}
P=P_{\mu\nu}P^{\mu\nu}=4L_F^2F=-\frac{2q^2}{r^4}\,,\label{P}
\end{eqnarray}
and
\begin{eqnarray}
r(P)=\left(\frac{2q^2}{-P}\right)^{1/4}\,.\label{rP}
\end{eqnarray}
Using \eqref{F} and \eqref{rP}, we find the function $F(P)$. According to \cite{Bronnikov}, the extremes of $F(P)$, $dF/dP=0$, play an important role in the description of electrically charged solutions. The maximums of the function $-F(-P)$ represent cusps in the representation $L(F)$ and how many are there, each cusp generates a new branch of the function $L(F)$. The minimums, being $ -F (-P_ {min}) = 0 $, we have a smooth branch change in $L(F)$. There are cases in which a local minimum of $ -F (-P) $ is not null, being another cusp in the representation of the function $L(F)$. This will be clear soon, when we specify concrete examples. This does not happen for magnetically charged solutions, as we will see later.

To analyze the regularity of the spacetime, we need to calculate the Kretschmann scalar, $K=R^{\mu\nu\alpha\beta}R_{\mu\nu\alpha\beta}$, that may be written as
\begin{equation}
K=f''(r)^2+\frac{4 \left(r^2 f'(r)^2+(f(r)-1)^2\right)}{r^4}.
\end{equation}
If the Kretschmann scalar does not present divergences, the spacetime does not have curvature singularities \cite{Bronnikov:2012wsj}.

To obtain regular solutions, some energy conditions must be relaxed \cite{Zaslavskii}. To analyze that, we may identify the components of the stress-energy tensor as $T^0_{\ 0}=\rho$, $T^1_{\ 1}=-p_r$ and $T^2_{\ 2}=T^3_{\ 3}=-p_t$, where $\rho$ is the energy density, $p_r$ is the radial pressure and $p_t$ is the tangential pressure. From the Einstein equations, the fluid quantities may be written as
\begin{eqnarray}
\rho(r)&=&\frac{1-f(r)-rf'(r)}{\kappa^2 r^2}=-p_r(r),\\
p_t(r)&=&\frac{2f'(r)+rf''(r)}{2\kappa^2r}.
\end{eqnarray}
With that, the energy conditions are
\begin{eqnarray}
WEC_{1,2}(r)&=&NEC_{1,2}(r)=SEC_{1,2}(r)=\rho+p_{r,t}\geq 0,\nonumber\\\label{EC2}\\
SEC_3(r)&=&\rho+p_r+2p_t \geq 0,\label{EC1}\\
WEC_{3}(r)&=&DEC_{1}(r)=\rho \geq 0,\label{EC3}\\
DEC_{2,3}(r)&=&\rho-\left|p_{r,t}\right| \geq0.\label{EC4}
\end{eqnarray}
About the strong energy condition, we may consider the following theorem:

{\bf Theorem} {\it Given a spherically symmetric solution from the Einstein equations, whose the stress-energy tensor satisfies the condition $T^{0}_{\ 0}=T^{1}_{\ 1}$, the strong energy condition will be violated for regions where we have $\left\{f'(r)< 0,f''(r)< 0\right\}$}.

{\bf Proof} If the component $T_{\mu\nu}$ satisfies $T^{0}_{\ 0}=T^{1}_{\ 1}$, we have an equation of state $\rho=-p_r$ and then \eqref{EC1} will depend only on $p_t$. Since the $SEC_3$ depends only on the tangential pressure, the sign of $f'(r)$ and $f''(r)$ will determine if $SEC_3(r)$ is positive or negative. So that, in regions where $\left\{f'(r)< 0,f''(r)< 0\right\}$, $SEC_3(r)$ is also negative and then SEC is violated.

Actually, here is general proof to the violation of SEC for regular, static and spherically symmetrical solutions. In \cite{Zaslavskii}, Zaslavskii defines the Tolman mass as being
\begin{eqnarray}
m_T&&=\int^{r_f}_{r_i}\left[T^{0}_{\;\;0}-T^{1}_{\;\;1}-T^{2}_{\;\;2}-T^{3}_{\;\;3}\right]dr\nonumber\\
&&=\int^{r_f}_{r_i}\left[\rho+p_r+2p_t\right]dr\,,\label{Tolman}
\end{eqnarray}
Zaslavskii shows that Tolman mass is always negative for a region within the event horizon for regular, static and spherically symmetric solutions, thus violating SEC for that region. This result is still valid for multihorizons, as established in \cite{Zaslavskii} and verified later in our solutions.

We may also define
\begin{equation}
\omega_r=\frac{p_r}{\rho}, \ \omega_t=\frac{p_t}{\rho}, \ \frac{\omega_t}{\omega_r}=\frac{p_t}{p_r}.
\end{equation}
If $\rho\geq 0$ then $DEC_1$, $WEC_1$ are satisfied and we can check the other energy conditions just by analyzing $\omega_r$ and $\omega_t$. Considering $\rho \geq0$, if $\omega_r> 1$ we affirm that $DEC_2$ is violated, and $\omega_t> 1$ we have that $DEC_3$ is violated. If $\omega_r <1$ then $NEC_1$, $WEC_1$, $SEC_1$ are violated and for $\omega_t <1$ we have that $NEC_2$, $WEC_2$, $SEC_2$ are violated. If $\omega_t <0$, then $SEC_3$ is not satisfied. 

\subsection{Balart-Vagenas solution}

Before building multihorizon solutions, let us look at an example of a regular solution.

A regular black hole model was proposed by Balart and Vagenas \cite{Balart}. They consider the mass function
\begin{equation}
M(r)=m\left(1+\frac{q^2}{4\beta m r}\right)^{-2\beta},
\end{equation}
where $\beta\geq3/2$ to guarantee the regularity and $\beta\leq3/2$ to satisfy the weak energy condition (WEC). If we choose $\beta=3/2$, we get
\begin{equation}
f(r)=1-\frac{432m^4r^2}{\left(q^2+6mr\right)^3}.\label{BVS}
\end{equation}
This solution has an event and a Cauchy horizon. If we expand $f(r)$ far from the event horizon and near to the black hole center, we find
\begin{eqnarray}
f(r)&\approx& 1-\frac{2m}{r}+\frac{q^2}{r^2}+O\left(\frac{1}{r^3}\right)\ \left(r\rightarrow \infty\right),\\
f(r)&\approx& 1-\frac{432m^4r^2}{q^6}+ O\left(r^3\right)\ \left(r\rightarrow 0\right).
\end{eqnarray}
So, for regions far from the event horizon, we find that the solution behaves like a Reissner-Nordström solution and near the center we see the behavior of a de Sitter solution.

The Kretschmann scalar is given by
\begin{eqnarray}
K(r)&=&\frac{4478976 m^8}{\left(6 m r+q^2\right)^{10}} \left(648 m^4 r^4-216 m^3 q^2 r^3\right.\nonumber\\
&+&\left.126 m^2 q^4 r^2+q^8\right),
\end{eqnarray}
which is regular for all values of $r$. The asymptotic behavior of this scalar is given by
\begin{eqnarray}
K(r\rightarrow 0)&&\sim \frac{4478976m^8}{q^{12}}+O\left(r\right)\,,\\
K(r\rightarrow \infty)&&\sim \frac{48m^2}{r^6}+O\left(r^{-7}\right).
\end{eqnarray}
We then see clearly that it is always regular with a constant curvature in the black hole center and the solution is asymptotically flat.

\begin{figure}[hbtp]
	\centering
	\includegraphics[width=\columnwidth]{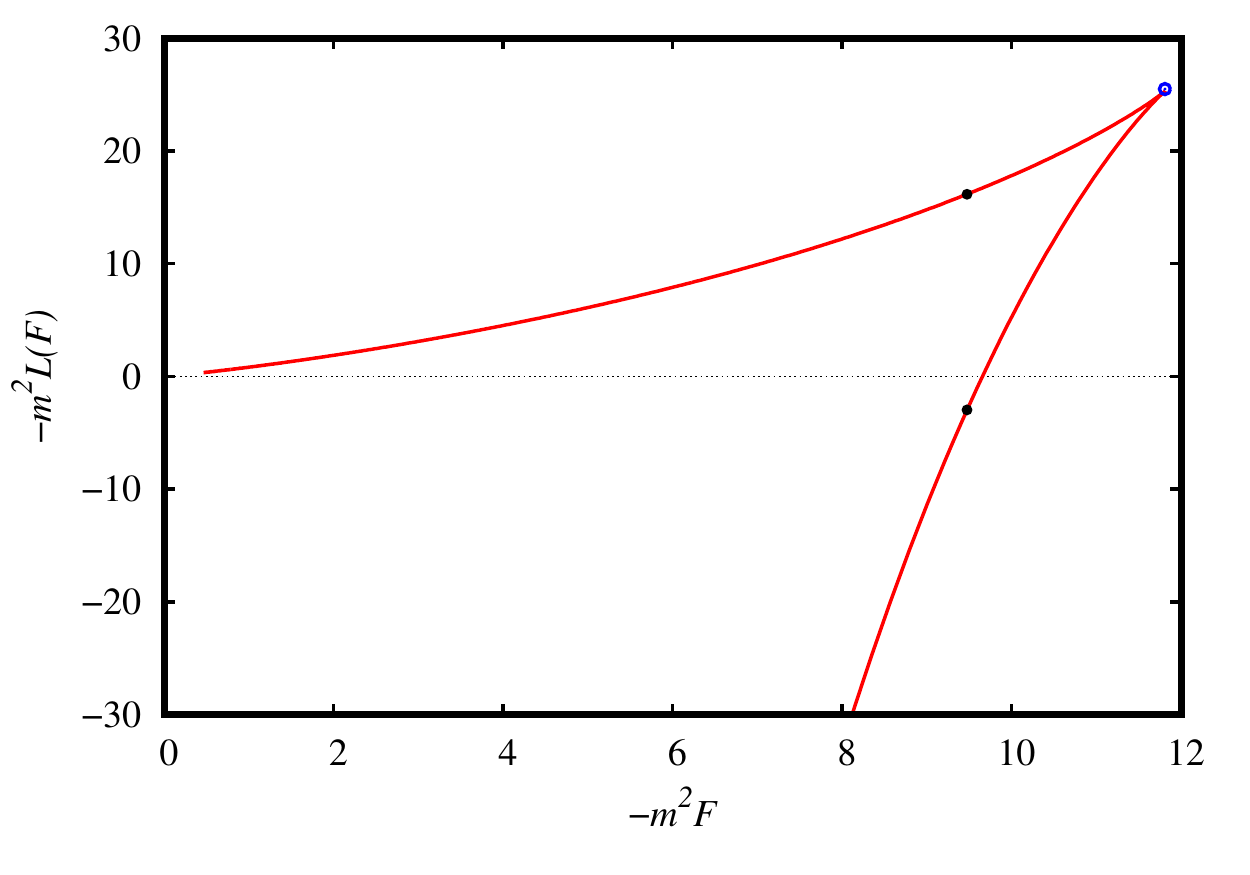}
	\caption{Graphical representation of $-L(F)\times -F$ with $q=0.8m$.}
	\label{LxF1}
\end{figure}

From \eqref{lel} and \eqref{lf}, with \eqref{ef}, the electromagnetic Lagrangian of the theory, that generates this solution, and the electric field are
\begin{eqnarray}
L(r)&=&\frac{1296m^4q^2\left(q^2-6mr\right)}{\left(q^2+6mr\right)^5},\label{L1}\\
F^{10}(r)&=&\frac{15552m^5q r^3}{\left(q^2+6mr\right)^5}.
\end{eqnarray}
The electric field is always regular and tends to zero at the infinity in the origin of the radial coordinate. Since we have the electric field, it is possible to construct the scalar $F$, that is given by
\begin{equation}
F(r)=-\frac{120932352m^{10}q^2r^6}{\left(q^2+6mr\right)^{10}}.\label{F1}
\end{equation}
We have the following asymptotic limits to $L(r)$ and $F(r)$
\begin{eqnarray}
L(r\rightarrow \infty)&&\sim -\frac{q^2}{ r^4}+O\left(r^{-5}\right), \\
L(r\rightarrow 0)&&\sim  \frac{1296 m^4}{ q^6}-\frac{46656 m^5 r}{ q^8}+O\left(r^2\right), \\
F(r\rightarrow \infty)&&\sim   -\frac{2q^2}{r^4}+O\left(r^{-5}\right), \\
F(r\rightarrow 0)&&\sim   -\frac{120932352 m^{10} r^6}{q^{18}}+O\left(r^7\right)\,.
\end{eqnarray}
So we have
\begin{eqnarray}
L(F)&&\hspace{-0.3cm}\sim F,r\rightarrow \infty\,, \\
L(F)&&\hspace{-0.3cm}\sim  \frac{1296 m^4}{ q^6}-\frac{1296\sqrt[6]{2} \sqrt[3]{3} \sqrt[6]{-\text{F}} m^{10/3}}{q^5},r\rightarrow 0.
\end{eqnarray}
Using \eqref{rP} and \eqref{L1}, we get
\begin{eqnarray}
-F(-P)=\frac{61917364224 m^{10} P \left(\frac{q^2}{P}\right)^{5/2}}{\left(12 m \sqrt[4]{\frac{q^2}{P}}+2^{3/4} q^2\right)^{10}}\,,
\end{eqnarray}
which has a maximum at $P=512 m^4/q^6$. As said before, a maximum of $-F(-P)$ represents a cusp in $L(F)$. We can parametrically represent the function $-L(-F)$ using \eqref{L1} and \eqref{F1}, whose the behavior is represented in Fig. \ref{LxF1}.

We see a cusp at $m^2F\approx-11.8098$, with two branch. This is a characteristic of regular electrically charged solutions. Branches appear because a single value of $F(r)$ corresponds to two positive real values of the radial coordinate $r$, resulting in two distinct values of $L(r)$. Let's look at a numerical example. To $m^2F(r)\approx-9.4748$, using \eqref{F1} to find the values of $r$, we have $r_1=0.10471057627514681m$ and $r_2=0.24750771046605197m$. These two values of $r$ give us two values of $L(r)$, which are $m^2L_1=m^2L(r_1)=2.96672$ and $m^2L_2=m^2L(r_2)=-16.1742$. This can now be verified in the graph of $-L(-F)$ in Fig. \ref{LxF1}.

From the stress-energy tensor, we get
\begin{eqnarray}
\rho(r)&=&\frac{1296 m^4 q^2}{\left(6 m r+q^2\right)^4}\,,\,p_r=-\rho(r)\\
p_t(r)&=&\omega_t(r)\rho(r)\,,\,\omega_t(r)=\frac{6 m r-q^2}{6 m r+q^2}.
\end{eqnarray}
We may write $p_{t}=p_{t}(\rho)$ inverting $\rho(r)$ and finding $r(\rho)$. With this, we have
\begin{equation}
p_t(\rho)=\omega_t(\rho) \rho, \ \omega_t(\rho)=\frac{36 m^2 \sqrt[4]{q^2}-(6 m+1) q^2 \sqrt[4]{\rho}}{36 m^2 \sqrt[4]{q^2}+(1-6 m) q^2 \sqrt[4]{\rho}}.
\end{equation}

\begin{figure}[hbtp]
	\centering
	\includegraphics[width=\columnwidth]{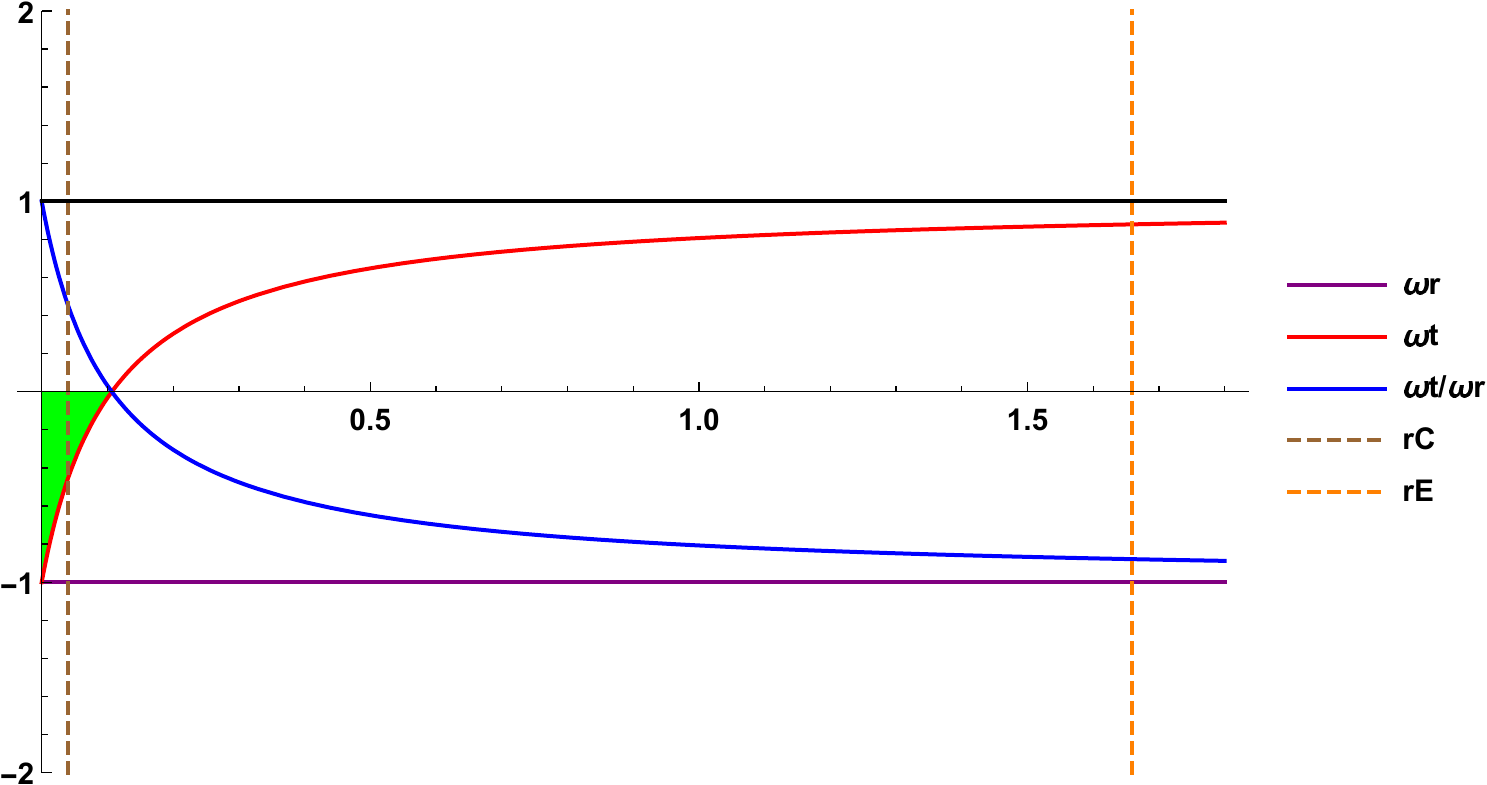} 
	\caption{Graphical representation of $\omega_{r}$, $\omega_{t}$ and $\omega_{r}/\omega_{t}$ as a function of the radial coordinate to $q=0.8m$. We see that $SEC_3$ is violated, green area.}
	\label{Omega_Balart_1}
\end{figure}

The solution behaves like an anisotropic fluid, $p_r\neq p_t$, with an equation of state $p_r=-\rho$. We have the following asymptotic limits for tangential pressure $p_t(r\rightarrow \infty)\sim\rho $ and $p_t(r\rightarrow 0)\sim -\rho $, and
\begin{eqnarray}
&&\rho(r\rightarrow \infty)\sim \frac{q^2}{8 \pi  r^4}+O\left(r^{-5}\right) ,\\
&&\rho(r\rightarrow 0)\sim\frac{162 m^4}{\pi  q^6}-\frac{3888 m^5 r}{\pi  q^8}+O\left(r^2\right).
\end{eqnarray}
The energy conditions are
\begin{eqnarray} \label{SEC_Balart}
SEC_{3}(r)&=&\frac{2592 m^4 q^2 \left(6 m r-q^2\right)}{\left(6 m r+q^2\right)^5},\\
WEC_1(r)&=&DEC_2(r)=0,\\
WEC_2(r)&=&\frac{15552 m^5 q^2 r}{\left(6 m r+q^2\right)^5},\\
WEC_{3}(r)&=&DEC_{1}(r)=\frac{1296 m^4 q^2}{\left(6 m r+q^2\right)^4},\\
DEC_3(r)&=&\frac{162 m^4 \left(-\left| q^4-6 m q^2 r\right| +6 m q^2 r+q^4\right)}{\pi  \left(6 m r+q^2\right)^5}.\nonumber\\
\end{eqnarray}
We can see that $SEC_3$ assumes negative values, and soon SEC is violated. In Fig. \ref{Omega_Balart_1} we see that to $q=0.8m$ we have a Cauchy horizon at $r=r_C=0.03952m$ and an event horizon at $r=r_E=1.65898m$. SEC is violated in the interval $r=[0,0.106667m]$. SEC is violated inside or outside the Cauchy horizon depending on the charge, but always inside the region bounded by the event horizon, while the other energy conditions are satisfied both inside and outside the event horizon.

\section{Multihorizon solutions}\label{sec3}

\subsection{Black holes with multihorizons}

The most simple black hole solution is described by the Schwarzschild metric, where the metric coefficient $g_{00}$ is
\begin{equation}
f(r)=1-\frac{2m}{r}.
\end{equation}
We may also write $f(r)$ as
\begin{equation}
f(r)=g_1(r)\left(r-r_1\right),
\end{equation}
with $g_1(r)=1/r$ and $r_1=2m$. To the Reissner-Nordström solution, when we have two horizons, $f(r)$ may be written as
\begin{equation}
f(r) = g_2(r)\left(r-r_1\right)\left(r-r_2\right),
\end{equation}
where $r_1$ and $r_2$ are the event horizon and Cauchy horizon radius, respectively. To the regular solution presented before, the metric coefficient is similar to Reissner-Nordström. If we consider the presence of cosmological constant, when we have a de Sitter-type solution, a new horizon appears in the solution and $f(r)$ becomes
\begin{equation}
f(r) =g_3(r) \left(r-r_1\right)\left(r-r_2\right)\left(r-r_3\right),
\end{equation} 
with $r_3$ being the radius of the cosmological horizon. The number of horizons is not limited to only three, in fact, it is possible to build structures with multiple horizons.

The metric coefficient to a multihorizon black hole may be written as 
\begin{eqnarray}
f(r)=&&g_N(r)\left(1-\frac{r_1}{r}\right)\left(1-\frac{r_2}{r}\right)\left(1-\frac{r_3}{r}\right)\times\nonumber\\
&&\times\dots\times\left(1-\frac{r_N}{r}\right),
\end{eqnarray}
where $r_i$, with $i=1,2,3,...,N$, represents the radius of each horizon, $g_N(r)$ is finite and $\lim_{r\rightarrow\infty}g_N(r)=1$. Depending of $g_N(r)$, the solution may be regular or singular in the center. However, we will write the metric coefficient in a different form in the next subsection. In \cite{Gao}, the authors analyzed the properties of a multihorizon solution, where they consider $g_N(r)=1$. This type of solution presents many horizons but only one singularity. The curvature invariants to this solution are
\begin{eqnarray}
R(r)&=&\frac{d^2}{dr^2}\left(\prod_{i=1}^{N} \left(1-\frac{r_i}{r}\right)\right)+\frac{4}{r} \frac{d}{dr}\left(\prod _{i=1}^{N}
\left(1-\frac{r_i}{r}\right)\right)\nonumber\\
&+&\frac{2}{r^2} \left(\prod_{i=1}^{N} \left(1-\frac{r_i}{r}\right)-1\right),\\
K(r)&=&\left(\hspace{-0.1cm}\frac{d^2}{dr^2}\hspace{-0.1cm}\left(\prod _{i=1}^{N} \hspace{-0.1cm}\left(1-\frac{r_i}{r}\right)\hspace{-0.1cm}\right)\hspace{-0.1cm}\right)^2\hspace{-0.1cm}+\hspace{-0.1cm}\frac{4}{r^4}\hspace{-0.1cm} \left(\hspace{-0.1cm}\left(\prod _{i=1}^{N}\hspace{-0.1cm} \left(1-\frac{r_i}{r}\right)\hspace{-0.1cm}-\hspace{-0.1cm}1\hspace{-0.1cm}\right)^2\right.\nonumber\\
&+&\left.r^2 \left(\frac{d }{d r}\left(\prod
	_{i=1}^{N} \left(1-\frac{r_i}{r}\right)\right)\right)^2\right).
\end{eqnarray}
The curvature scalar is null to one and two horizons and is singular in the black hole center to $N\geq 3$, while the Kretschmann scalar is singular in $r=0$ to all values of $N$. With some modifications, we may also construct regular multihorizon solutions in the General Relativity.

\subsection{Regular black holes with multihorizons}

To regular solutions we write the coefficient $g_{00}$ as
\begin{equation}
f(r)=\prod_{i=1}^{N}\left(1-\frac{2M_i(r)}{r}\right),
\label{fgeral}
\end{equation}
where $\lim\limits_{r\rightarrow \infty}M_i(r)/r=0$. Which mass function tends to a constant in the infinity, $\lim\limits_{r\rightarrow \infty}M_i(r)=m_i$, where the ADM mass is the sum of these constants, $m_{ADM}=\sum_{i=1}^{N}m_i$. One way to satisfy these conditions is to use the product of known regular solutions. The solution obtained from known regular solutions is also regular and has the number of horizons equal to the sum of the number of horizons of the solutions that compose it.

\subsubsection{First example with four horizons}

Let us consider a solution with the metric coefficient
\begin{equation}
f(r)=\left(1-\frac{2M_1(r)}{r}\right)\left(1-\frac{2M_2(r)}{r}\right),\label{fs}
\end{equation}
where
\begin{equation}
M_1(r)=\frac{mr^3}{\left(r^2+q^2\right)^{3/2}}\ \mbox{and} \ M_2(r)=me^{-q^2/2mr}. 
\end{equation}
We considerer $m_1=m_2=m$. The functions $M_1(r)$ and $M_2(r)$ were proposed by James Bardeen \cite{BardeenConf} and Hristu Culetu \cite{Culetu}, respectively. The solution proposed by Bardeen violates the strong and dominant energy condition while the Culetu solution violates the strong and weak energy condition. The solution \eqref{fs} is asymptotically flat, has a de Sitter core. To $q\leq q_{ext}^{BD}= 4m/(3\sqrt{3})$ presents four horizons, when $q_{ext}^{BD}<q<q_{ext}^{CL}$ two horizons and one horizon to $q=q_{ext}^{CL}$, where $q_{ext}^{CL}=\frac{2m}{\sqrt{e}}$ is the extreme charge to the Culetu solution. 

\begin{figure}[htb!]
	\includegraphics[width=\columnwidth]{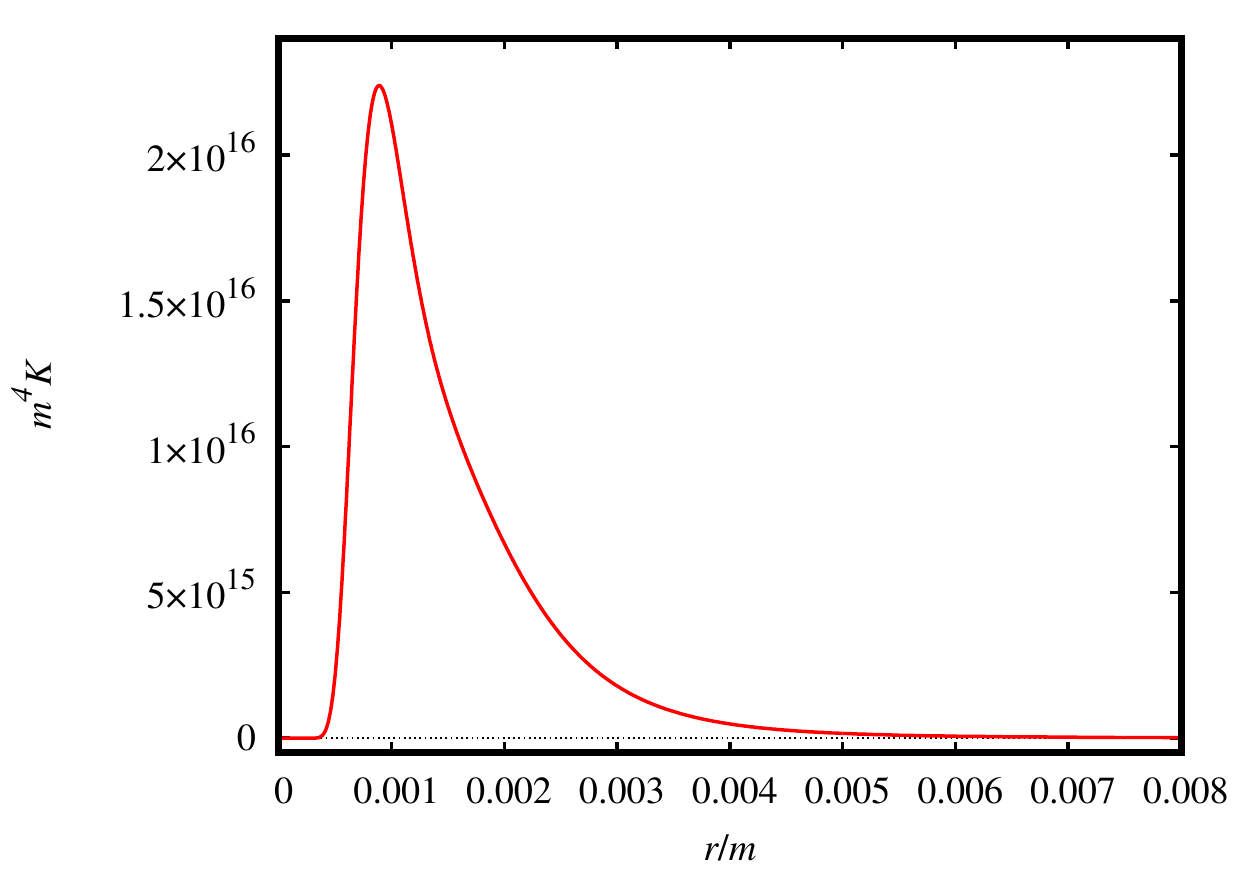}
	\caption{Kretschmann scalar to \eqref{fs} with $q=0.1m$}
	\label{inv1}
\end{figure}

The asymptotic behavior of the Kretschmann scalar is given by
\begin{eqnarray}
&K&\hspace{-0.1cm}\sim \frac{96m^2}{q^6}+ e^{-\frac{q^2}{m r}} \left(\frac{q^8}{4 m^2 r^{10}}-\frac{4 q^6}{m r^9}+O\left(r^{-8}\right)\right)\nonumber\\
&+&\hspace{-0.3cm}e^{-\frac{q^2}{2 m r}} \left(\frac{4 \sqrt{q^2}}{r^5}-\frac{32 m}{\sqrt{q^2}
   r^4}+O\left(r^{-3}\right)\right),r\rightarrow 0.
\end{eqnarray}
From the curvature invariant, Fig. \ref{inv1}, the regularity of the spacetime is highlighted.

\begin{figure*}[htb!]
	\includegraphics[width=\columnwidth]{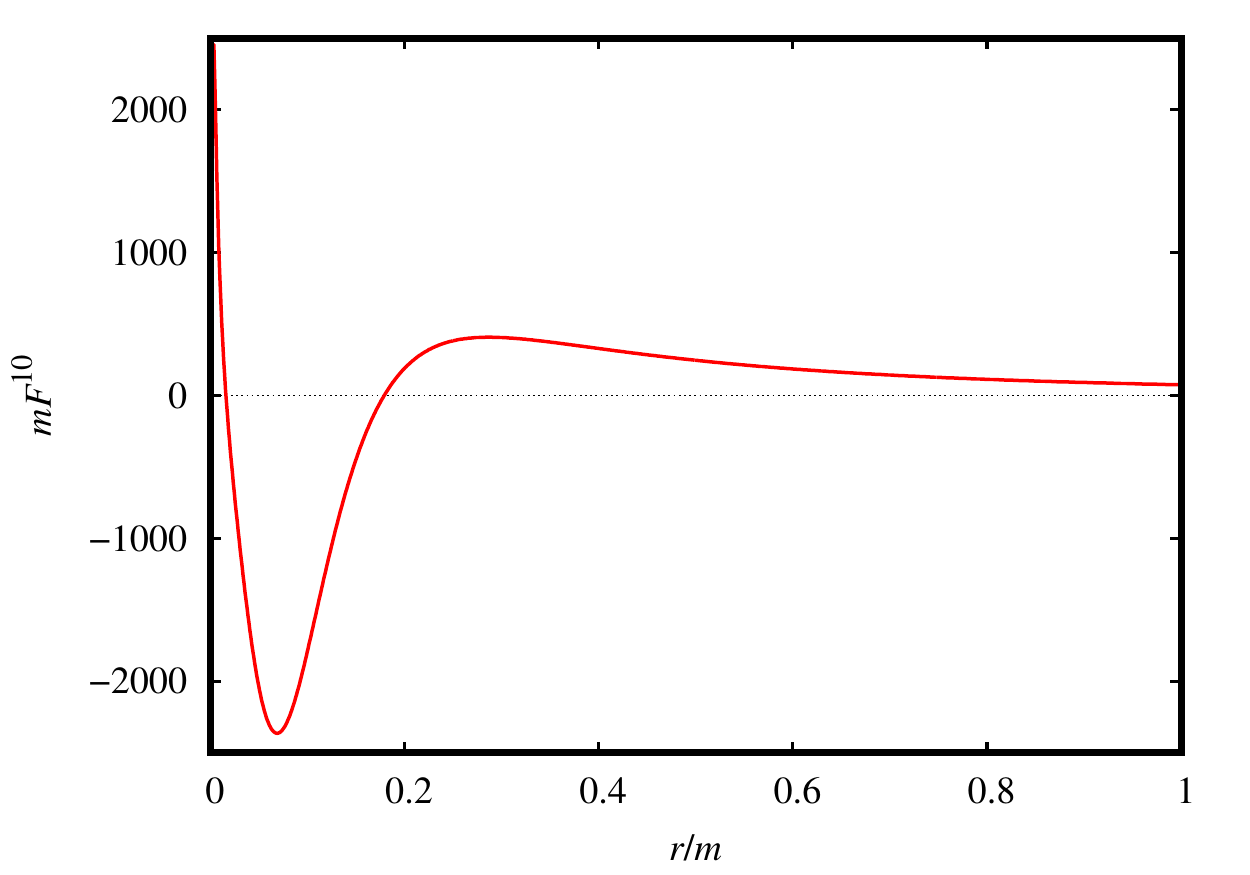}
	\includegraphics[width=\columnwidth]{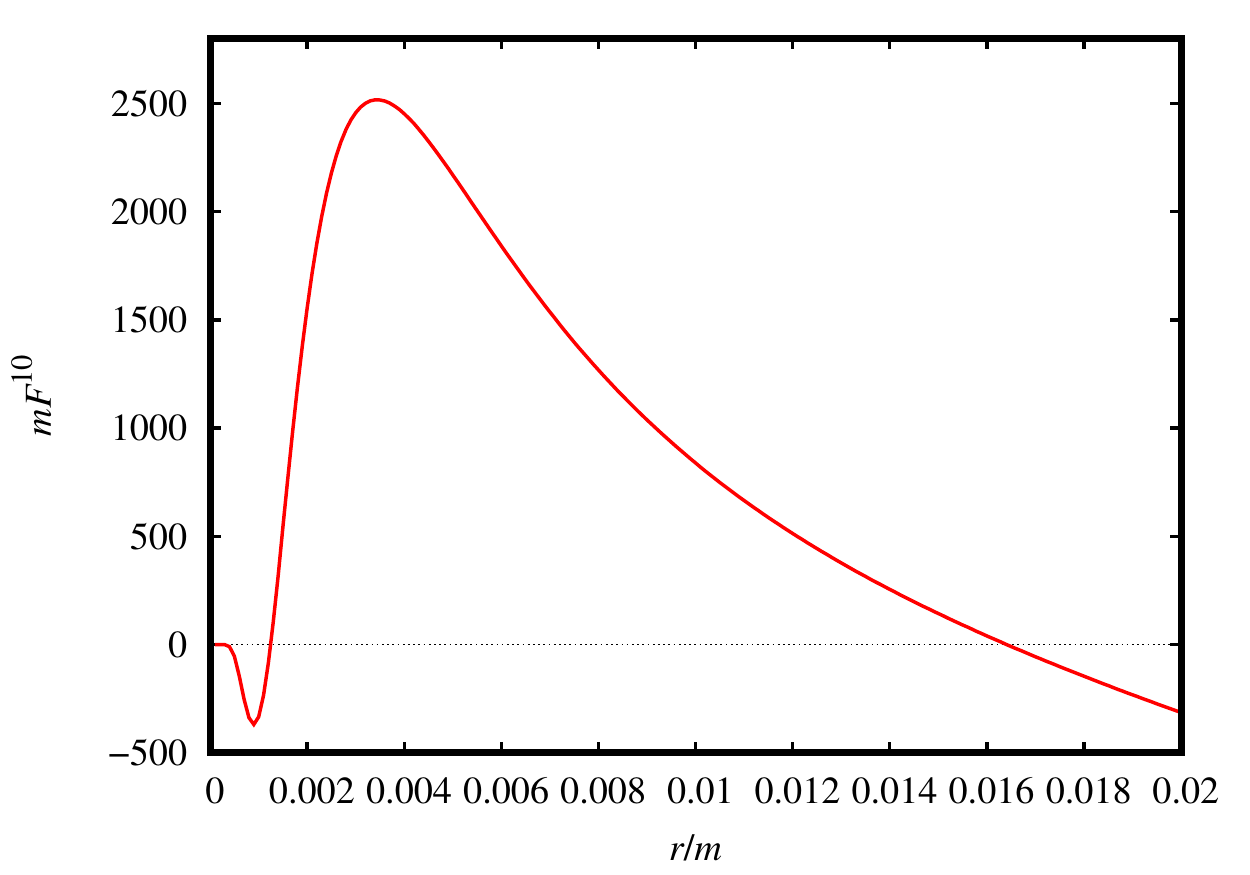}
	\caption{Electric field associated to the solution with four horizons to $q=0.1m$.}
	\label{F10S1}
\end{figure*}

In relation to the electromagnetic sector, the intensity of the electric field is shown in Fig. \ref{F10S1}. The electric field goes to zero in the black hole center and in the infinity. As the sing of $F^{10}$ changes it means that the field will repel and attract the same test particle for different regions. 

\begin{figure*}
\centering
\subfigure[]{\label{F21}\includegraphics[width=\columnwidth]{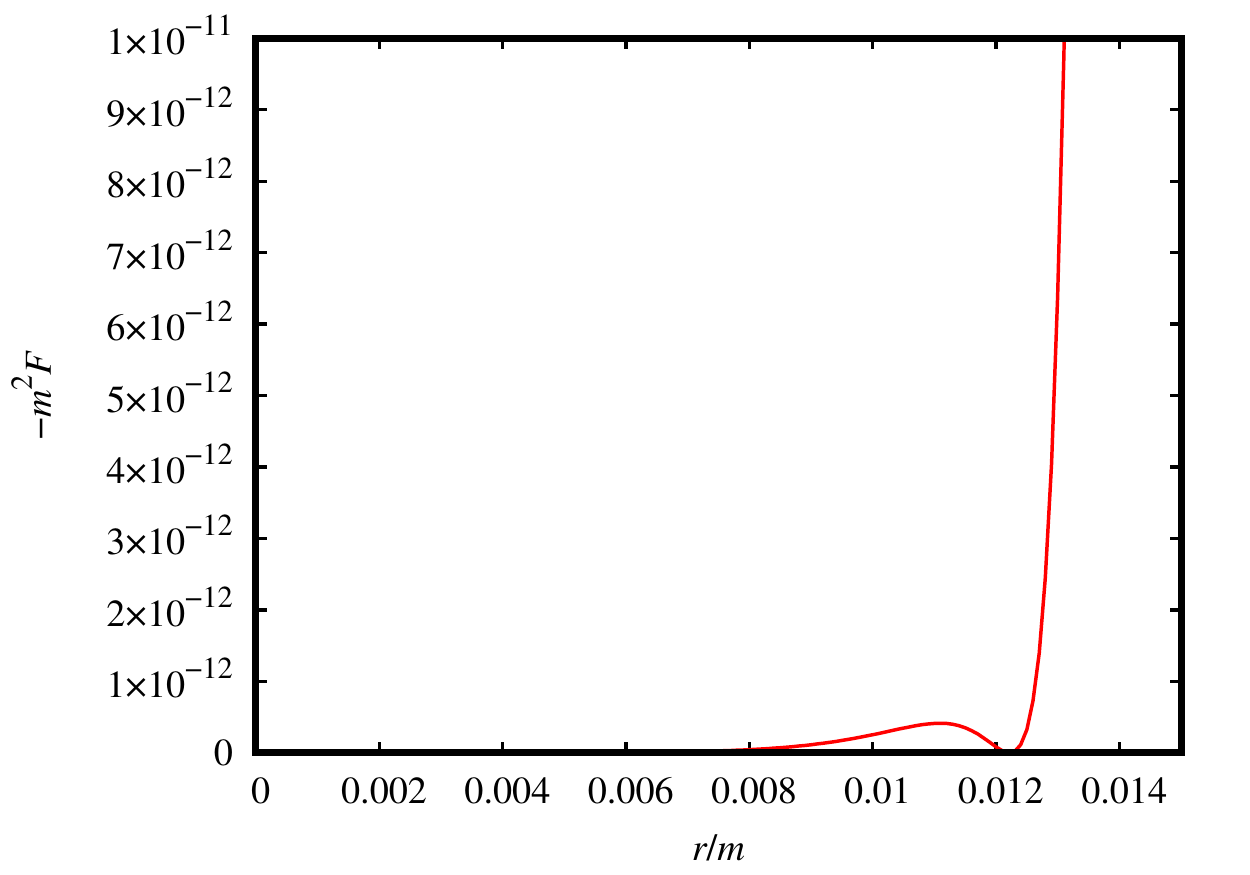}}
\subfigure[]{\label{F22}\includegraphics[width=\columnwidth]{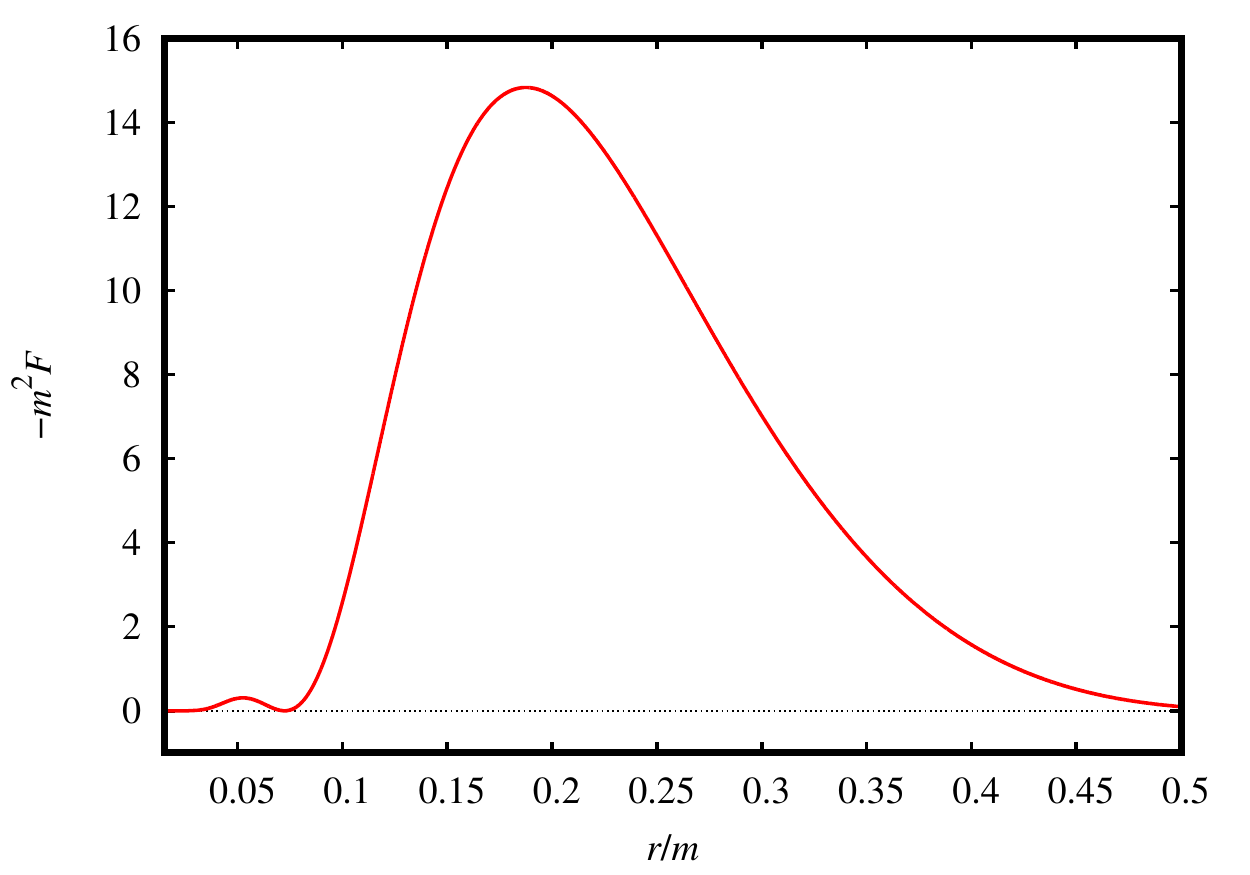}}
\subfigure[]{\label{F23}\includegraphics[width=\columnwidth]{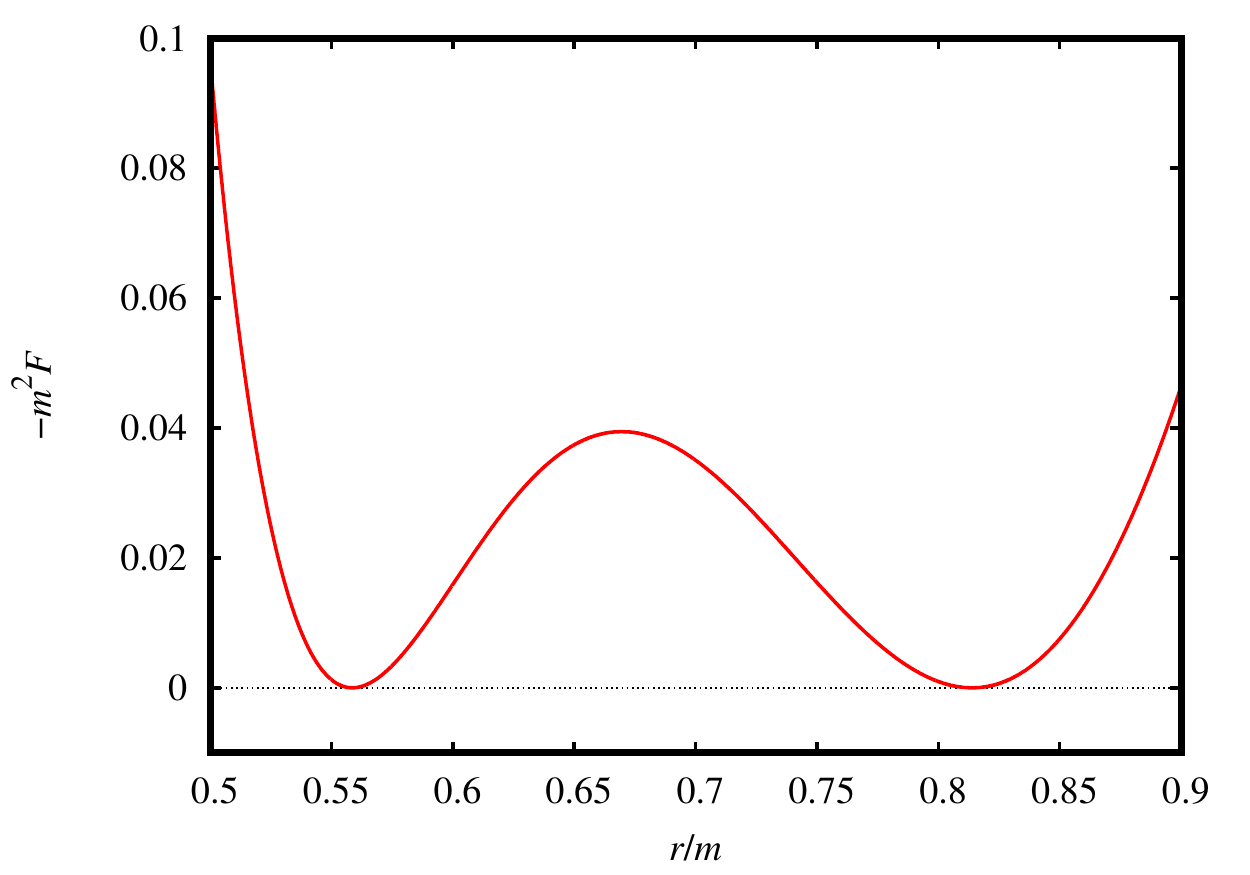}}
\subfigure[]{\label{F24}\includegraphics[width=\columnwidth]{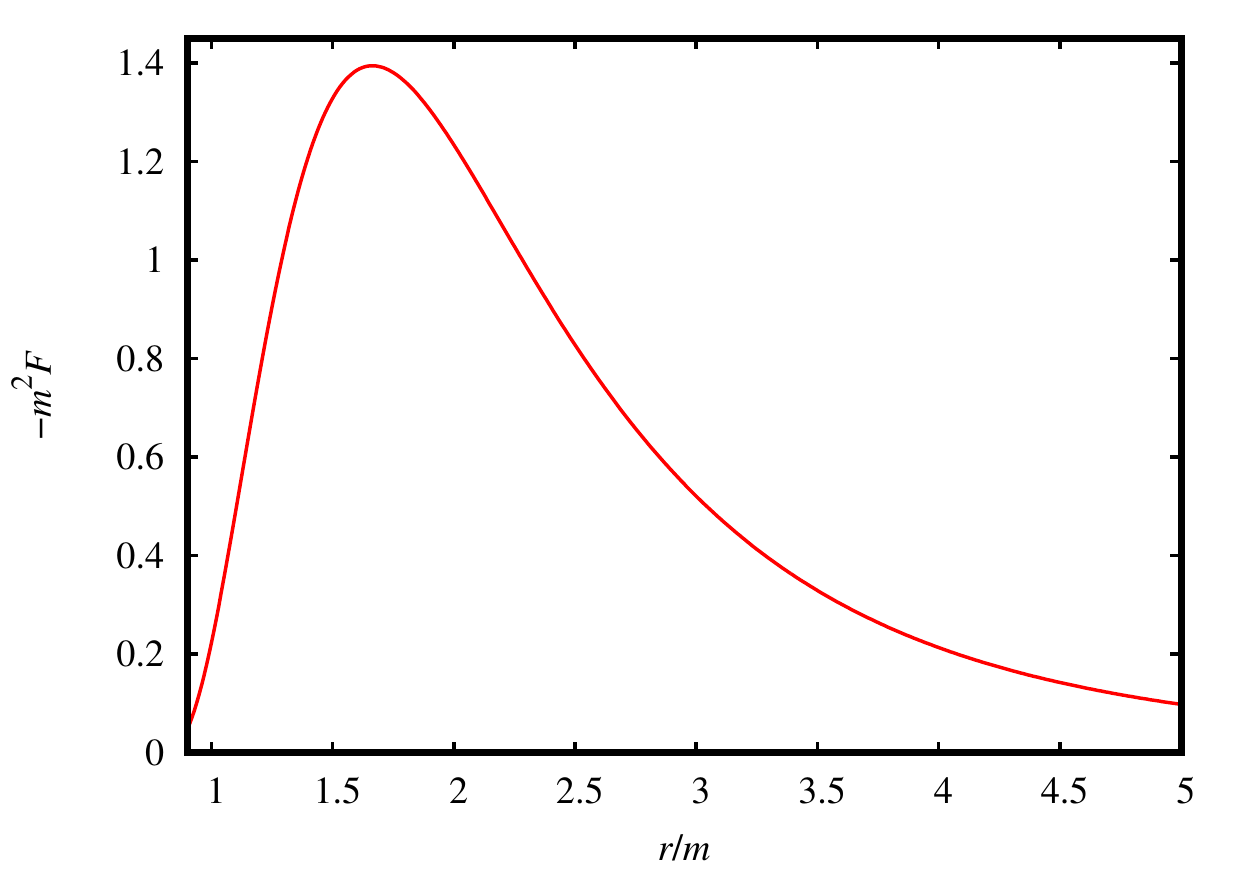}}
\caption{Behavior of $-F(r)$ to $q=0.769m$. The radial coordinate range is (a) $[0,0.015m]$, (b) $(0.015m,0.5m]$, (c) $(0.5m,0.9m]$ and (d) $(0.9m,5m]$.}
\label{FS1}
\end{figure*}

\begin{figure*}
\centering
\subfigure[]{\label{LxF21}\includegraphics[width=\columnwidth]{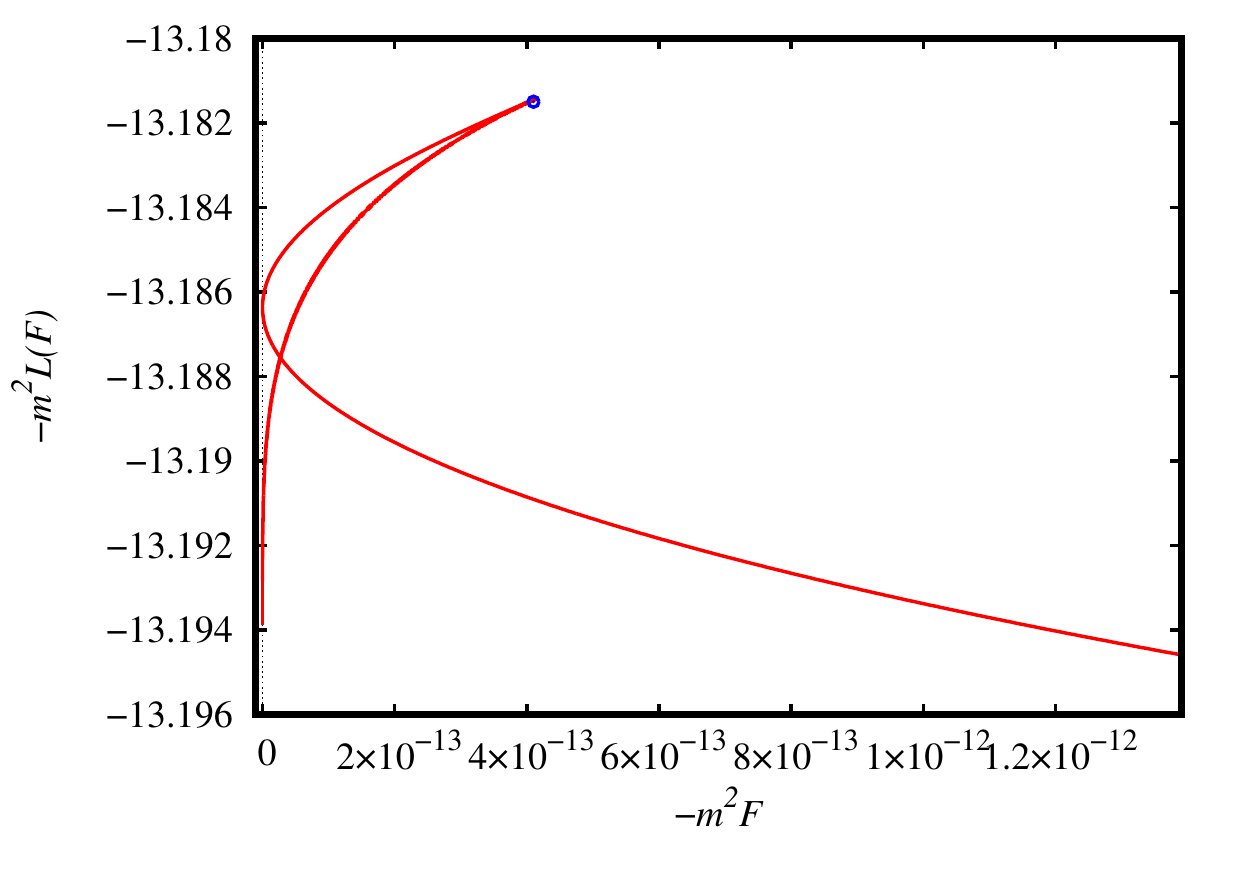}}
\subfigure[]{\label{LxF22}\includegraphics[width=\columnwidth]{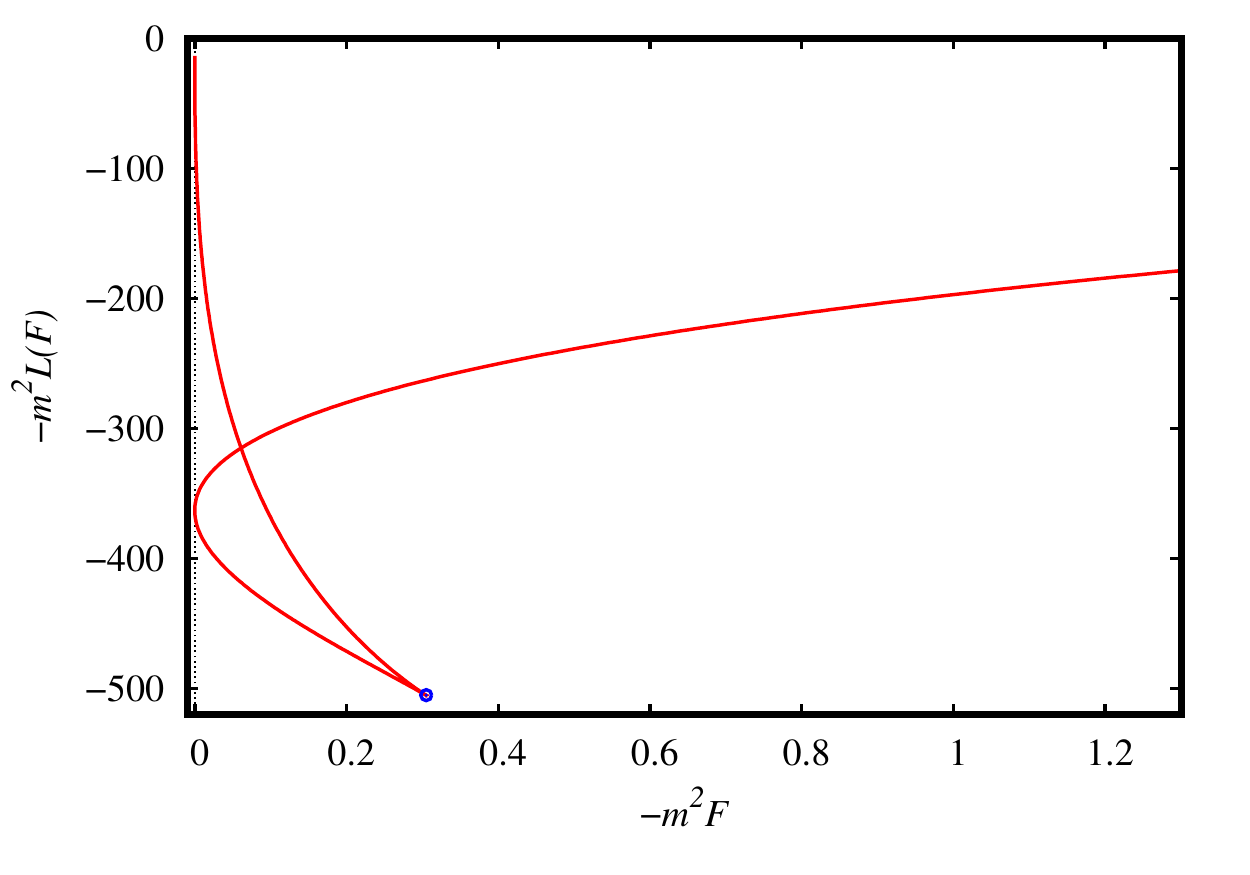}}\\
\subfigure[]{\label{LxF23}\includegraphics[width=\columnwidth]{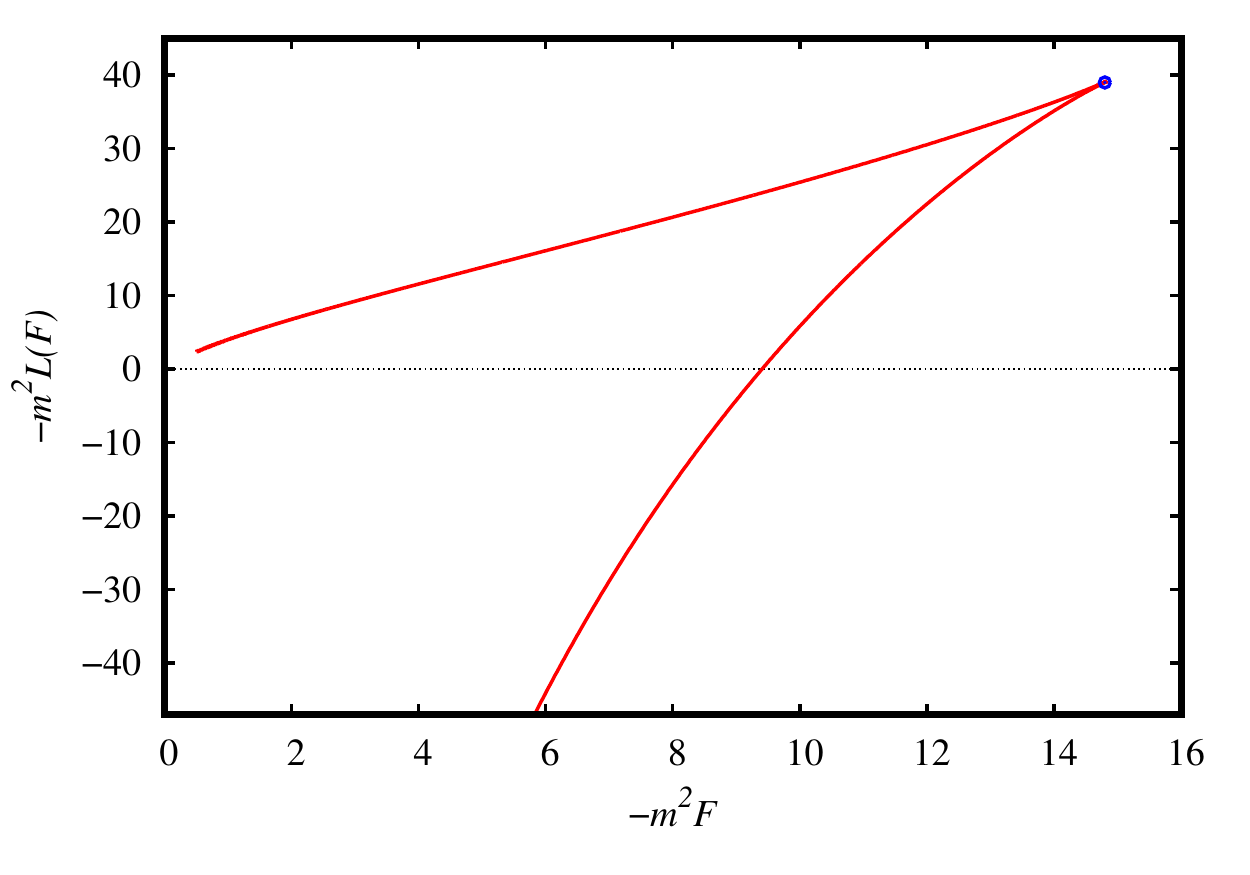}}
\subfigure[]{\label{LxF24}\includegraphics[width=\columnwidth]{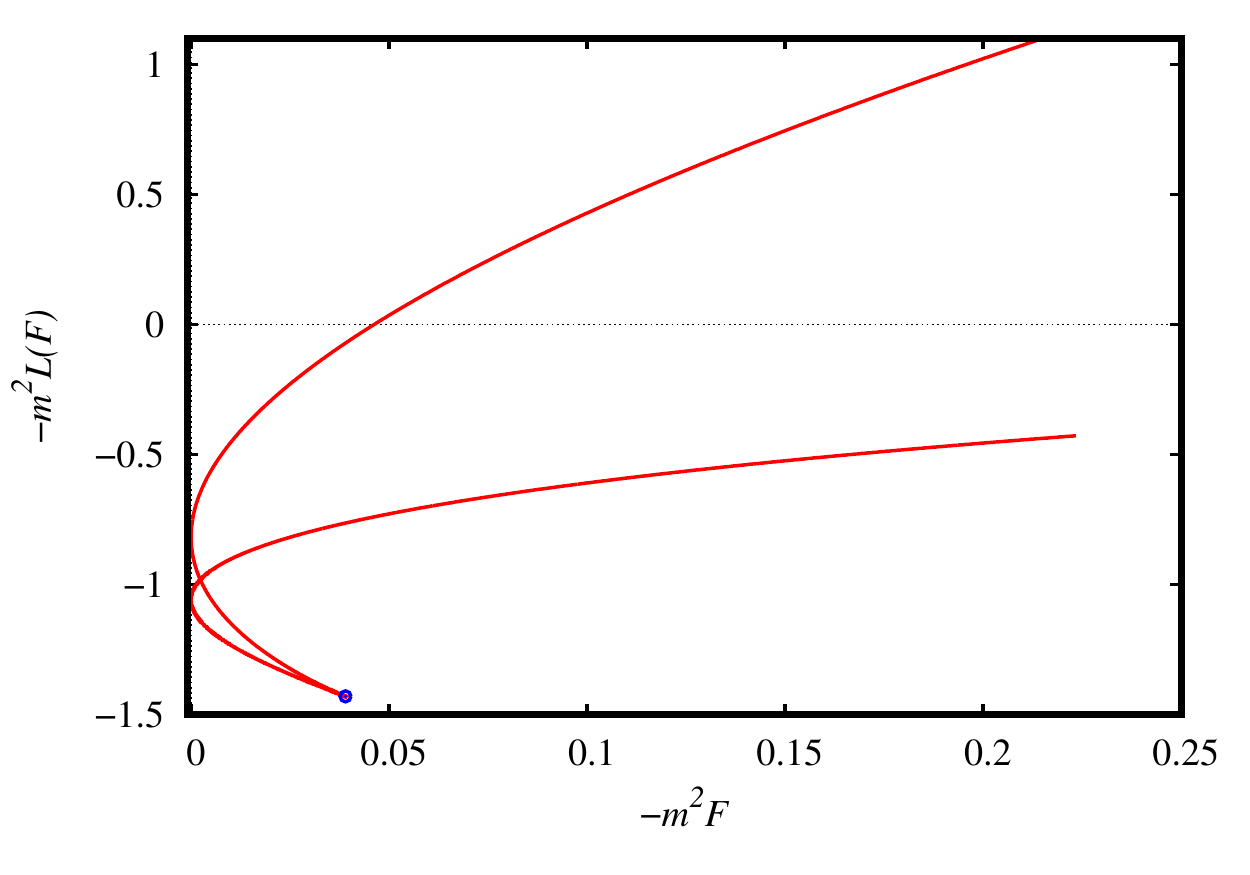}}\\
\subfigure[]{\label{LxF25}\includegraphics[width=\columnwidth]{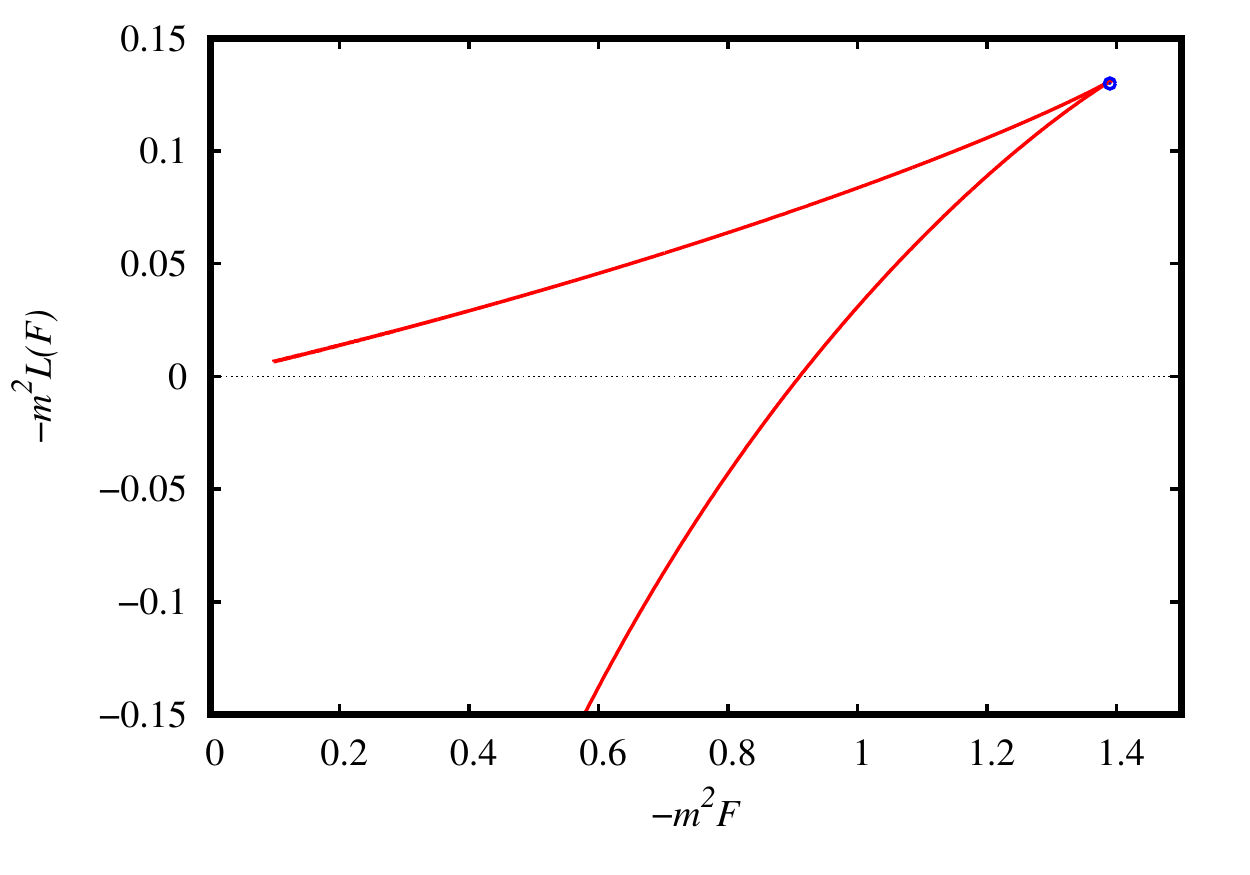}}
\caption{Parametric representation of $-L(-F)$ to $q=0.769m$. The radial coordinate range is (a) $[0,0.014m]$, (b) $(0.014m,0.1m]$, (c) $(0.1m,0.45m]$, (d) $(0.45m,m]$ and (e) $(m,5m]$.}
\label{LxF2}
\end{figure*}

The asymptotic behavior of $L(r)$ and $F(r)$ is given by
\begin{eqnarray}
&&L(r\rightarrow \infty)\sim  -\frac{4 m^2+q^2}{  r^4}+O\left(r^{-5}\right),\\
&&L(r\rightarrow 0)\sim   \frac{6 m \sqrt{q^2}}{ q^4}+\frac{q^4 e^{-\frac{q^2}{2 m r}}}{4 m r^5},\\
&&F(r\rightarrow \infty)\sim  -\frac{2\left(4m^2+q^2\right)^2}{q^2 r^4} +O\left(r^{-5}\right), \\
&&F(r\rightarrow 0)\sim -\frac{225 m^2 r^8}{2 q^{12}}.
\end{eqnarray}
So we have
\begin{eqnarray}
L(F)&\sim &  F,r\rightarrow \infty,\label{LF11}\\
L(F)&\sim & \frac{6 m \sqrt{q^2}}{ q^4} +\frac{15 \sqrt[4]{15m} e^{-\frac{\sqrt[4]{15} \sqrt{q}}{2 \sqrt[8]{-2F} m^{3/4}}}}{4\ (-2F)^{5/8} q^{7/2}},r\rightarrow 0.\nonumber\\
\end{eqnarray}
We note that close to the center, the Lagrangian does not behave like Maxwell, only to $r\rightarrow \infty$. Replacing \eqref{fs} in \eqref{lel}-\eqref{F} and \eqref{ef}, we get $L(r)$ and $F(r)$. In Fig. \ref{FS1}, we see the behavior of the scalar $F$ as a function of $r$. Replacing \eqref{rP} in $F(r)$ we have the analytical expression to $F(P)$. We will not show $-F(-P)$ because it is extensive, numerically we can find nine extreme to this function of which five local maximums and four local minimums, where $-F(-P_{min})=0$. We may also see that from Fig. \ref{FS1}, since $dF(r)/dr=0 \rightarrow dF(P)/dP=0$, it's due to the fact that $dF/dr=(dF/dP)(dP/dr)$ and $dP/dr$ is null only to $r\rightarrow \infty$. We have five cusps in the function $-L(-F)$, which we can see in the Fig. \ref{LxF2}. So we have ten distinct branches for $L(F)$.

From the components of the stress-energy we find
\begin{eqnarray}
\rho(r)&=&\frac{e^{-\frac{q^2}{2 m r}}}{8 \pi  r^4 \left(q^2+r^2\right)^{5/2}} \bigg(m^2 \left(4 r^5-8 q^2 r^3\right)\nonumber\\
&-&2 m q^2 r^2 \left(r^2 \left(1-3 e^{\frac{q^2}{2m r}}\right)+q^2\right)\nonumber\\
&+&q^2 \left(q^2+r^2\right)^{5/2}\bigg)=-p_r,\\
p_t(r)&=&\frac{e^{-\frac{q^2}{2 m r}}}{32 \pi  m r^5 \left(q^2+r^2\right)^{7/2}} \bigg(-2 q^2 r^2 \left(3 e^{\frac{q^2}{2 m r}}+1\right)\nonumber\\
&+&4 m^2 q^2 r^3 \left(r^4 \left(9	e^{\frac{q^2}{2 m r}}-4\right)+2 q^4\right)\nonumber\\
&+&2 m \left(2 q^2 r^3	\left(q^2+r^2\right)^{5/2}+q^4 r \left(q^2+r^2\right)^2 \times \right.\nonumber\\
&\times&\left.\left(2 \sqrt{q^2+r^2}+r\right)\right)+q^4\left(-\left(q^2+r^2\right)^{7/2}\right)\nonumber\\
&+&8 m^3 \left(2 q^4 r^4-11 q^2 r^6+2 r^8\right)\bigg).
\end{eqnarray}
The solution behaves like an anisotropic fluid with a de Sitter-type equation of state, $\rho=-p_r$. 
We have the following asymptotic limits
\begin{eqnarray}
&&\rho(r\rightarrow \infty)=-p_r(r\rightarrow \infty)\sim  \frac{4 m^2+q^2}{8 \pi  r^4}+O\left(r^{-5}\right)  ,\nonumber\\ \\ 
&&\rho(r\rightarrow 0)=-p_r(r\rightarrow 0)\sim e^{-\frac{q^2}{2 m r}} \left(\frac{q^2}{8 \pi  r^4}+ O\left(r^{-3}\right)\right)\nonumber\\
&&+\frac{6m}{8\pi\left(q^2\right)^{3/2}},\\
&&p_t(r\rightarrow \infty)\sim \frac{4 m^2+q^2}{8 \pi  r^4}+O\left(r^{-5}\right),\\
&&p_t(r\rightarrow 0)\sim -e^{-\frac{q^2}{2 m r}} \left(\frac{q^4}{32 \pi m r^5}+O\left(r^{-4}\right)\right)\nonumber\\
&&-\frac{6m}{8\pi\left(q^2\right)^{3/2}}\,.
\end{eqnarray}

\begin{figure}[hbtp]
	\centering
	\includegraphics[width=\columnwidth]{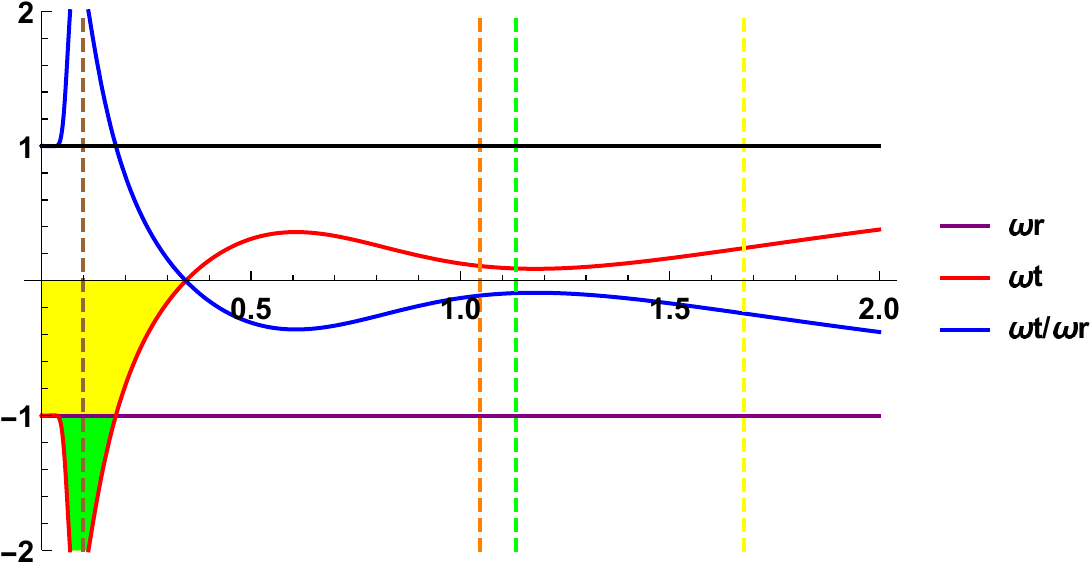}
	\caption{Graphical representation of $\omega$ in terms of the radial coordinate to $q=0.75m$.}
	\label{Omega_Bardeen-Culetu}
\end{figure}

The energy density is not always positive, however, we may impose some constraints on the charge to guarantee the positivity. To values of charge $q<0.721m$, we have an interval in which the energy density is negative, so we will choose $q$ such that the density is always greater than zero for all values of $r$, this is necessary to satisfy WEC and NEC. The function $WEC_1(r)$ is identically zero. To guarantee that some energy conditions will be always satisfied, we must impose some constraints on the charge. In order to determine which energy conditions are met in this case we look at Fig. \ref{Omega_Bardeen-Culetu}, we can see, for $q=0.75m $ the SEC is violated in the regions yellow and green, while DEC is violated in the green region. All energy conditions are met in the region outside the black hole.

\subsubsection{Second example with four horizons}

\begin{figure}
	\includegraphics[width=\columnwidth]{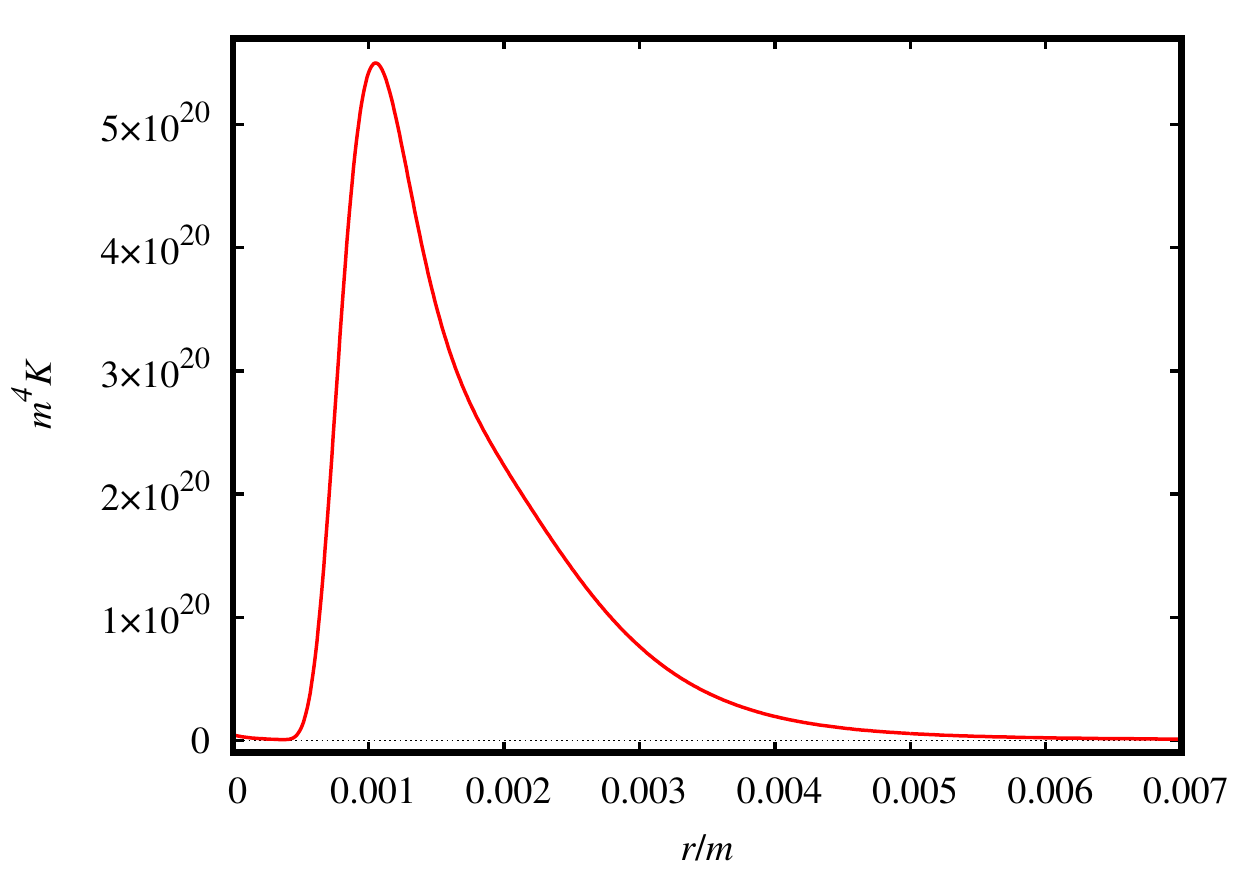}
	\caption{Behavior of the Kretschmann scalar to \eqref{SS} with $q=0.1m$.}
	\label{CIS2}
\end{figure}

Let's consider a solution like \eqref{fs}, however, we replace $M_1(r)$ by $M_3(r)$, where this mass function is the one that generates the solution \eqref{BVS}. So that, the metric coefficient $g_{00}$ is
\begin{equation}
f(r)=\left(1-\frac{432m^4r^2}{\left(q^2+6mr\right)^3}\right)\left(1-\frac{2me^{-q^2/2mr}}{r}\right).\label{SS}
\end{equation}
This solution is asymptotically flat, regular in all spacetime and behaves like de Sitter in the black hole center and has four horizons for $q<q_{ext}^{CL}$. The Kretschmann scalar is shown in Fig. \ref{CIS2} and we have no divergences, which implies in no curvature singularities. We have the following limits
\begin{eqnarray}
K(r\rightarrow \infty)&\sim&\frac{192}{r^6}+O(r^{-7}),\\
K(r\rightarrow 0)&\sim&\frac{4478976m^8}{q^{12}}+O(r).
\end{eqnarray}
So the Kretschmann scalar is regular in the center and in the infinity of the radial coordinate.

\begin{figure}[htb!]
	\includegraphics[width=\columnwidth]{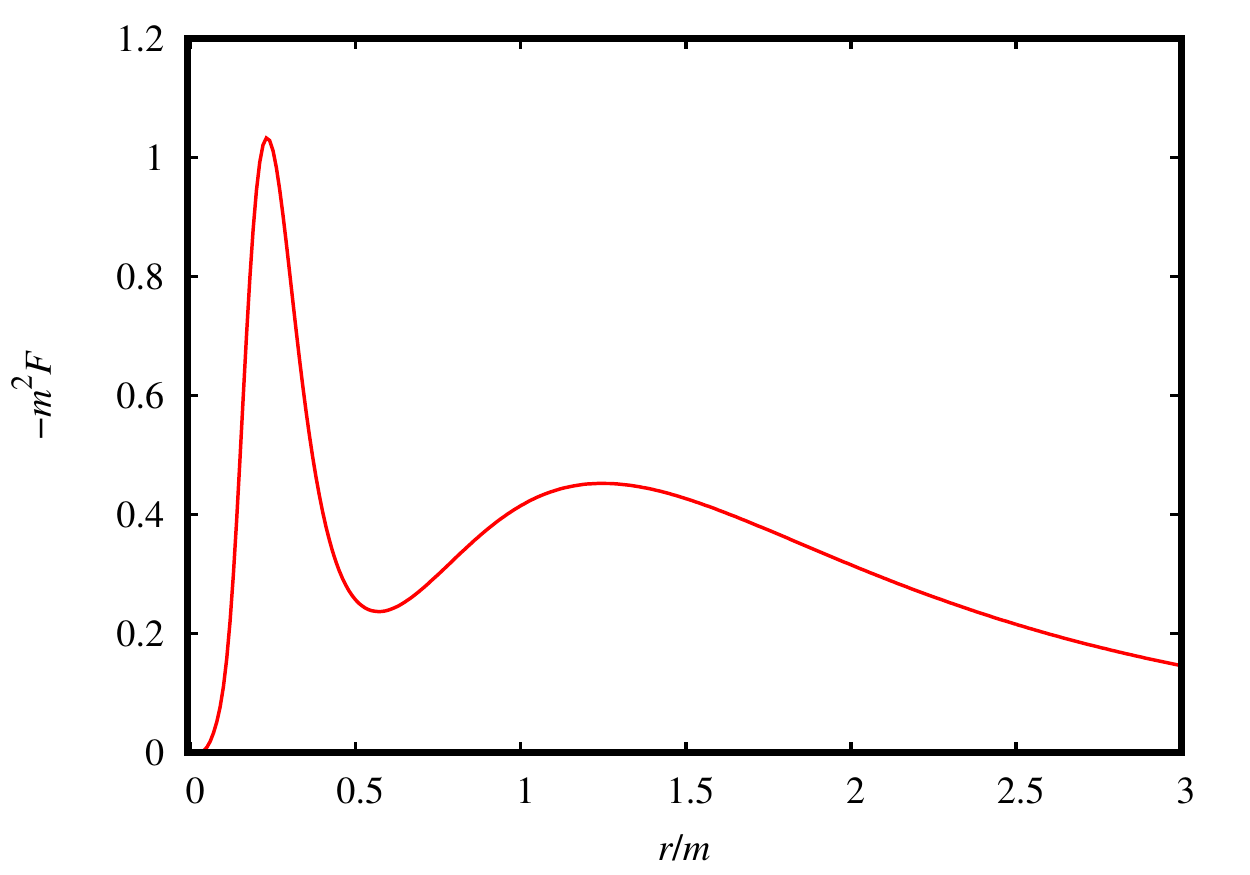}
	\caption{Graphical representation of $-F(r)$ to $q=1.21m$.}
	\label{Fr1}
\end{figure}

\begin{figure*}[htb!]
	\subfigure[]{\includegraphics[width=\columnwidth]{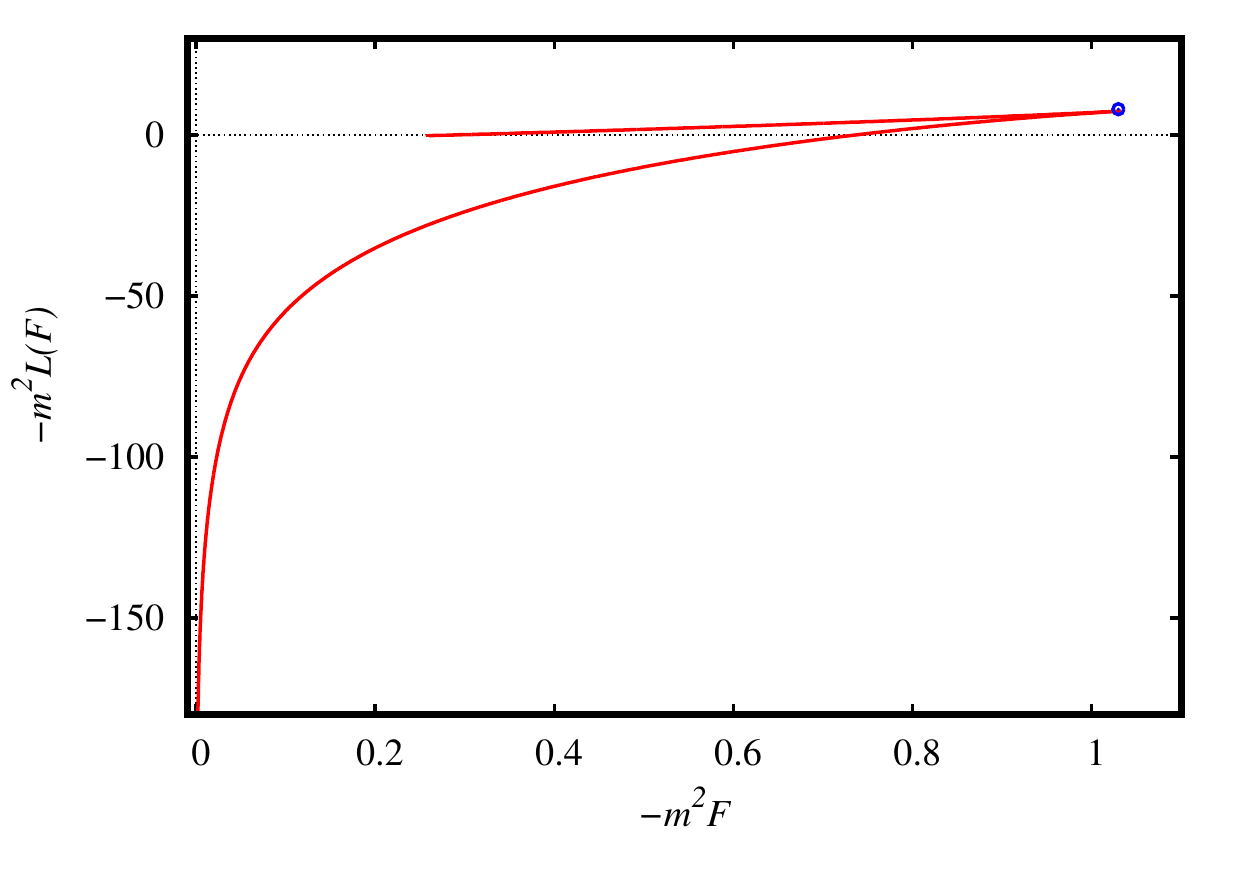}}
	\subfigure[]{\includegraphics[width=\columnwidth]{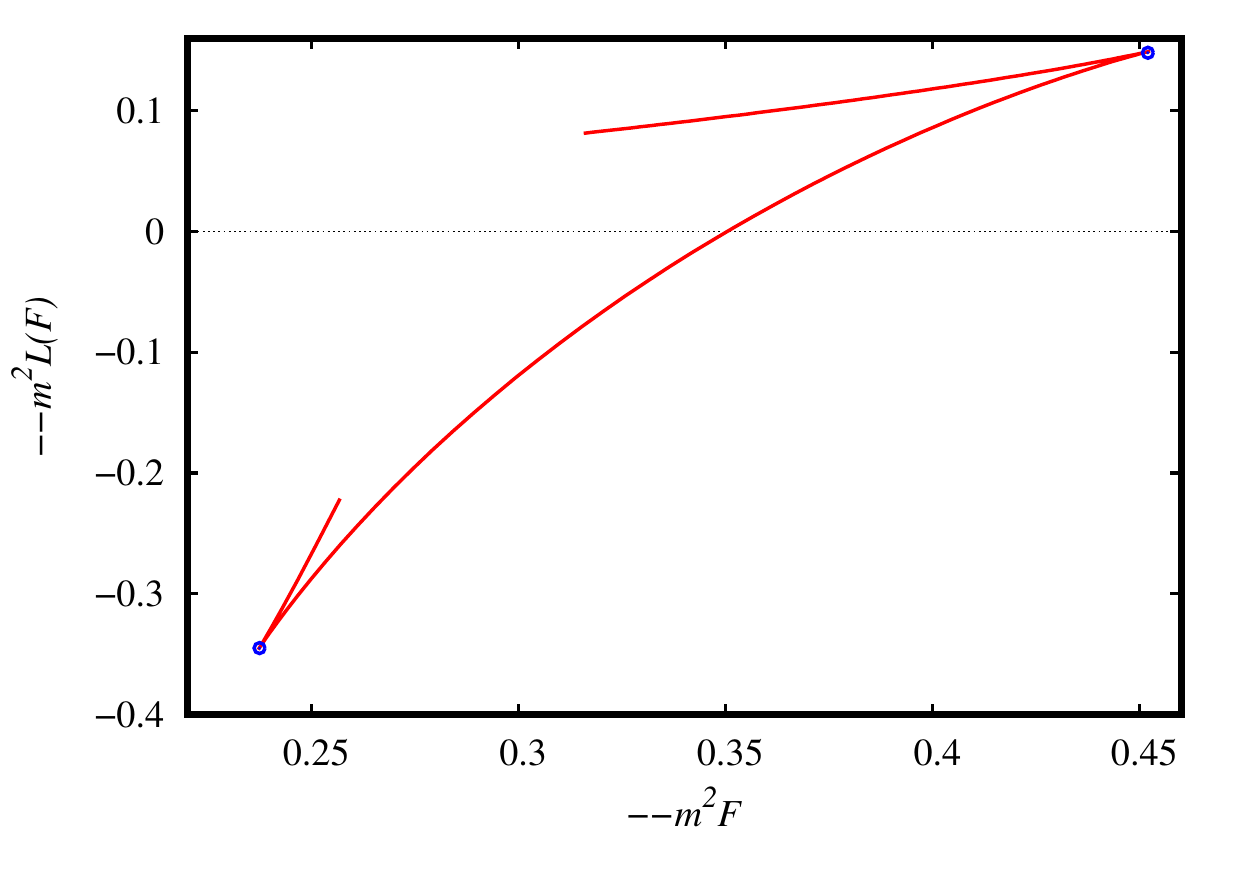}}
	\caption{Parametric representation of $-L(-F)$ to $q=1.21m$. The interval of $r$ is (a) $[0,0.5m]$ and (b) $(0.5m,2m]$.}
	\label{LxF3}
\end{figure*}

The asymptotic limits of the electromagnetic quantities are

\begin{eqnarray}
L(r\rightarrow \infty)&\sim &  -2\frac{2 m^2+q^2}{ r^4}+O(r^{-5}),\\
L(r\rightarrow 0)&\sim &  \frac{1296 m^4}{ q^6}+\frac{q^4 e^{-\frac{q^2}{2 m r}}}{4 m r^5},\\
F(r\rightarrow \infty)&\sim & -8\frac{\left(2m^2+q^2\right)^2}{q^2 r^4}+O(r^{-5}) ,\\
F(r\rightarrow 0)&\sim &- \frac{120932352m^{10}r^{6}}{q^{18}} .
\end{eqnarray}
So, the asymptotic dependence is the same of \eqref{LF11} to $r\rightarrow \infty$. We have, 
\begin{eqnarray}
L(F)&\sim& F,r\rightarrow \infty,\\
L(F)&\sim&  \frac{839808 \sqrt[6]{2} \sqrt[3]{3m^{10}}m^4 e^{-\frac{3\ 2^{5/6} 3^{2/3} m^{2/3}}{\sqrt[6]{-F} q}}}{q^{11}(-F)^{5/6}}\nonumber\\
&+& \frac{1296 m^4}{ q^6},r \rightarrow 0.
\end{eqnarray}
Solving $-F'(r)=0$ numerically, we have three extremes, two local maximums and a local minimum to $-F(r_{min})$ not null, which implies in a $-F(-P_{min})$ not null. This point in $-F(-P)$ represents a cusp in $-L(-F)$, this is a new result in the literature. We represent the graph of $-F(r)$ in Fig. \ref{Fr1}, where we may see that it has two maximums and one minimum as tends to zero to $r\rightarrow 0$ and $r\rightarrow \infty$. We represent the parametric graph of $-L(-F)$ in Fig. \ref{LxF3}, where we see three cusps and four branches. 

From the stress-energy tensor, we find
\begin{eqnarray}
\rho(r)&=&\frac{e^{-\frac{q^2}{2 m r}} }{\kappa^2  r^4 \left(6	m r+q^2\right)^4}\bigg(5184 m^6 r^4+24 m q^8 r+q^{10}\nonumber\\
&-&432 m^4 q^2 r^3 \left(10 m-3 r \left(e^{\frac{q^2}{2 mr}}+1\right)\right)\nonumber\\
&-&432 m^3 q^4 r^2 (m-2 r)+216 m^2 q^6 r^2\bigg)=-p_r(r),\nonumber\\\\
p_t(r)&=&\frac{e^{-\frac{q^2}{2 m r}} }{4\kappa^2 m r^5 \left(6 m r+q^2\right)^5}\bigg(124416 m^7 r^6 \bigg(5 m\nonumber\\
&-&r \left(e^{\frac{q^2}{2 mr}}+1\right)\bigg)-10368 m^6 q^2 r^5 \times\nonumber\\
&&\left(28 m+r-5 r e^{\frac{q^2}{2 m r}}\right)-864 m^5 q^4 r^4	\bigg(46 m\nonumber\\
&-&r \left(2 e^{\frac{q^2}{2 m r}}+41\right)\bigg)+144 m^4 q^6 r^3 (24 m+95 r)\nonumber\\
&+&48 m^3 q^8 r^2 (9m+35 r)-16 m^2 q^{10} r^2\nonumber\\
&-&18 m q^{12} r-q^{14}\bigg).
\end{eqnarray}
As the examples before we have the behavior of an anisotropic fluid. We have the following asymptotic limits
\begin{eqnarray}
\rho(r\rightarrow \infty)&\sim & \frac{2 m^2+q^2}{4 \pi  r^4}-\frac{q^2\left(48m+7q^2\right)}{48m\pi r^5}+O\left(r^{-6}\right),\nonumber\\
\\
\rho(r\rightarrow 0)&\sim &\left(\frac{162 m^4}{\pi  q^6}+O\left(r\right)\right)+\nonumber\\
&+&e^{-\frac{q^2}{2 m r}}\left(\frac{q^2}{8 \pi  r^4}+O\left(r^{-2}\right)\right),\\
p_t(r\rightarrow \infty)&\sim& \frac{2 m^2+q^2}{4 \pi  r^4}-\frac{q^2\left(48m+7q^2\right)}{32m\pi r^5}+O\left(r^{-6}\right),\nonumber\\ \\
p_t(r\rightarrow 0)&\sim &-\left(\frac{162 m^4}{\pi  q^6}+O\left(r\right)\right)+\nonumber\\
&-&e^{-\frac{q^2}{2 m r}}\left(\frac{q^4}{32 \pi  mr^5}+O\left(r^{-2}\right)\right),\\
\omega_t(r\rightarrow \infty)&\sim & 1- \frac{48 m^2 q^2+7 q^4}{24 m r \left(2 m^2+q^2\right)}+O\left(r^{-2}\right),\\
\omega_t(r\rightarrow 0)&\sim &-1. 
\end{eqnarray}

\begin{figure}[htb!]
	\includegraphics[width=\columnwidth]{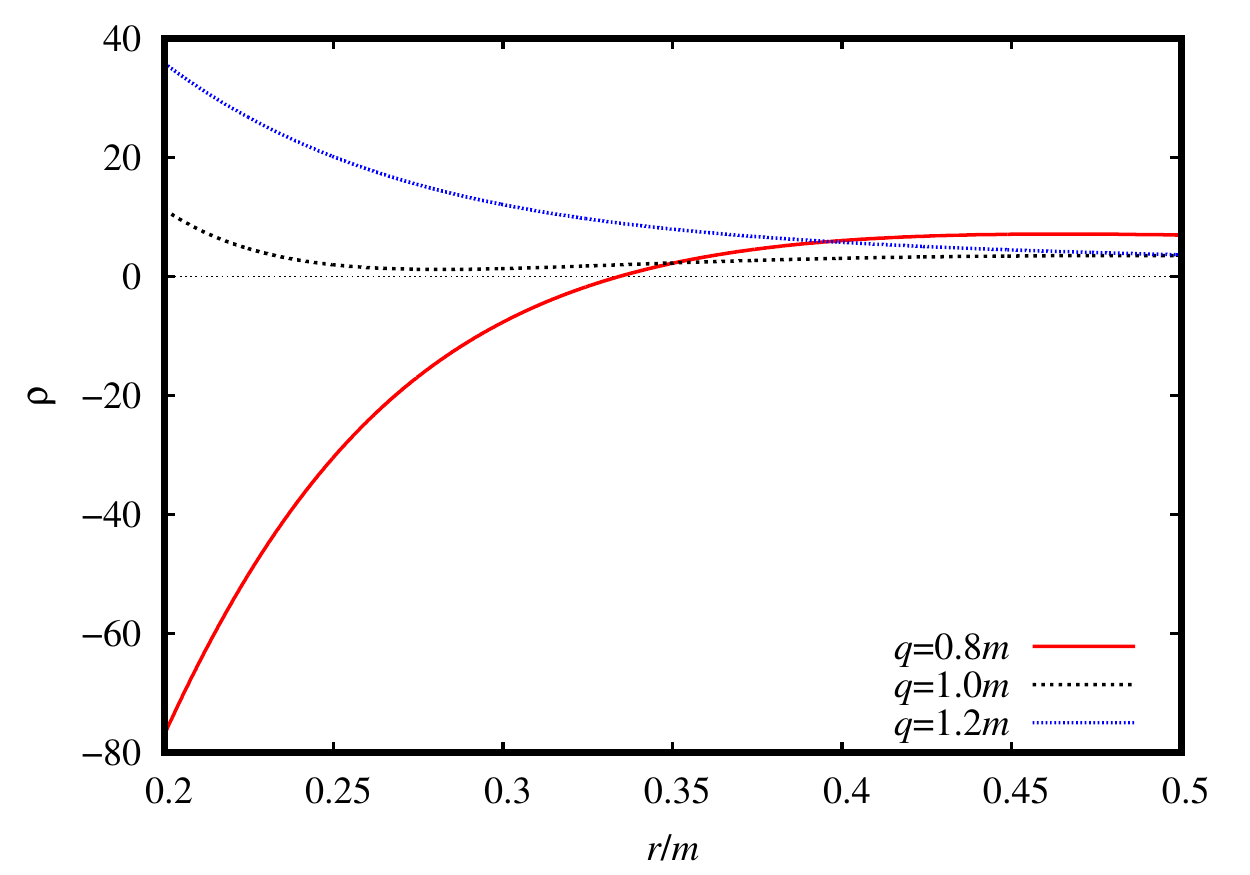}\vspace{-.1cm}
	\caption{Energy density to the solution \eqref{SS} for different values of charge.}
	\label{densSOL2}
\end{figure}

\begin{figure}[hbtp]
	\centering
	\includegraphics[width=\columnwidth]{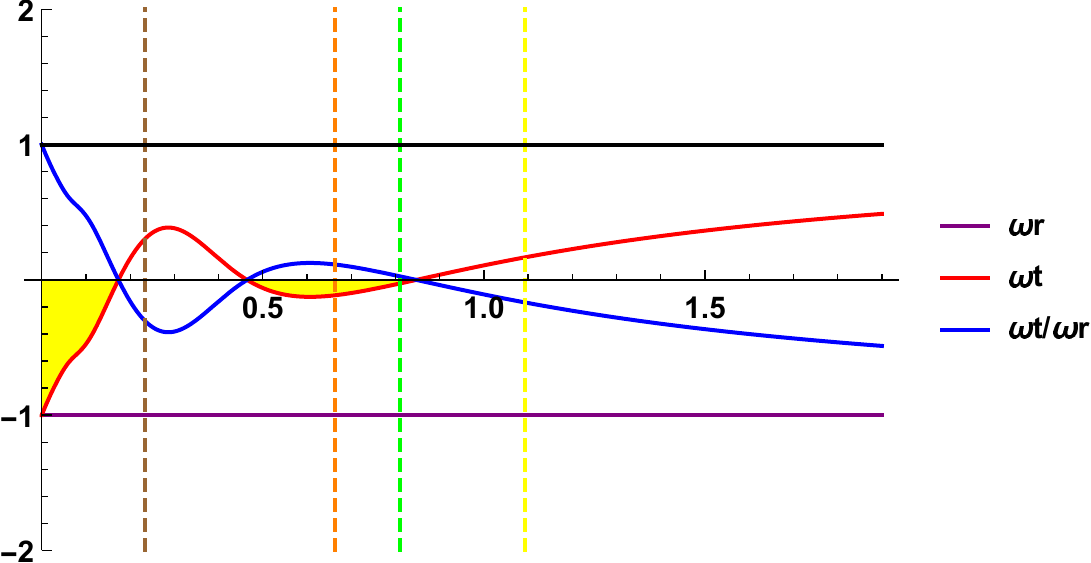}
	\caption{Graphical representation of $\omega$ as a function of the radial coordinate to $q=1.21m$. SEC is violated in the yellow region. Each horizontal line represents a horizon.}
	\label{Omega_Balart-Culetu}
\end{figure}

The energy density admits negative values for some ranges of $r$, however, if we impose constraints in the electric charge, it is possible to guarantee the positivity. In Fig. \ref{densSOL2}, we see that for some values of charge we may have positive energy density, so that, through Fig. \ref{Omega_Balart-Culetu} we see that SEC is violated in the yellow region which is inside the black hole. NEC, DEC and WEC are satisfied.

\subsubsection{Example with six horizons}

Let's consider the case where we combined the three functions previously. To this model, the metric coefficient $g_{00}$ is
\begin{eqnarray}
f(r)&=&\left(1-\frac{2M_2(r)}{r}\right)\left(1-\frac{2M_3(r)}{r}\right)\left(1-\frac{2M_4(r)}{r}\right),\nonumber\\
\label{TS}
\end{eqnarray}
where $M_4(r)$ is a mass function proposed by Irina Dymnikova \cite{Dymnikova}, that generates a regular solution and is given by
\begin{equation}
M_4(r)=\frac{2m\tan^{-1}\left(\frac{8mr}{\pi q^2}\right)}{\pi}-\frac{16m^2q^2 r}{64m^2r^2+\pi^2q^4}.
\end{equation}
The Dymnikova solution has a extreme given by $q^{DC}_{ext}=1.07304927103275m$. This solution has up to six horizons, however, the number of horizons decreases as the electrical charge increases. As the cases before, this spacetime is regular and asymptotically flat. Near to the center, the Kretschmann scalar behaves as
\begin{eqnarray}
K(r\rightarrow 0)&\sim & \frac{2048 \left(65536+41472 \pi ^4+6561 \pi ^8\right) m^8}{3 \pi ^8 q^{12}}\nonumber\\
&+&O(r).\nonumber\\
\end{eqnarray}

\begin{figure}[htb!]
	\includegraphics[width=\columnwidth]{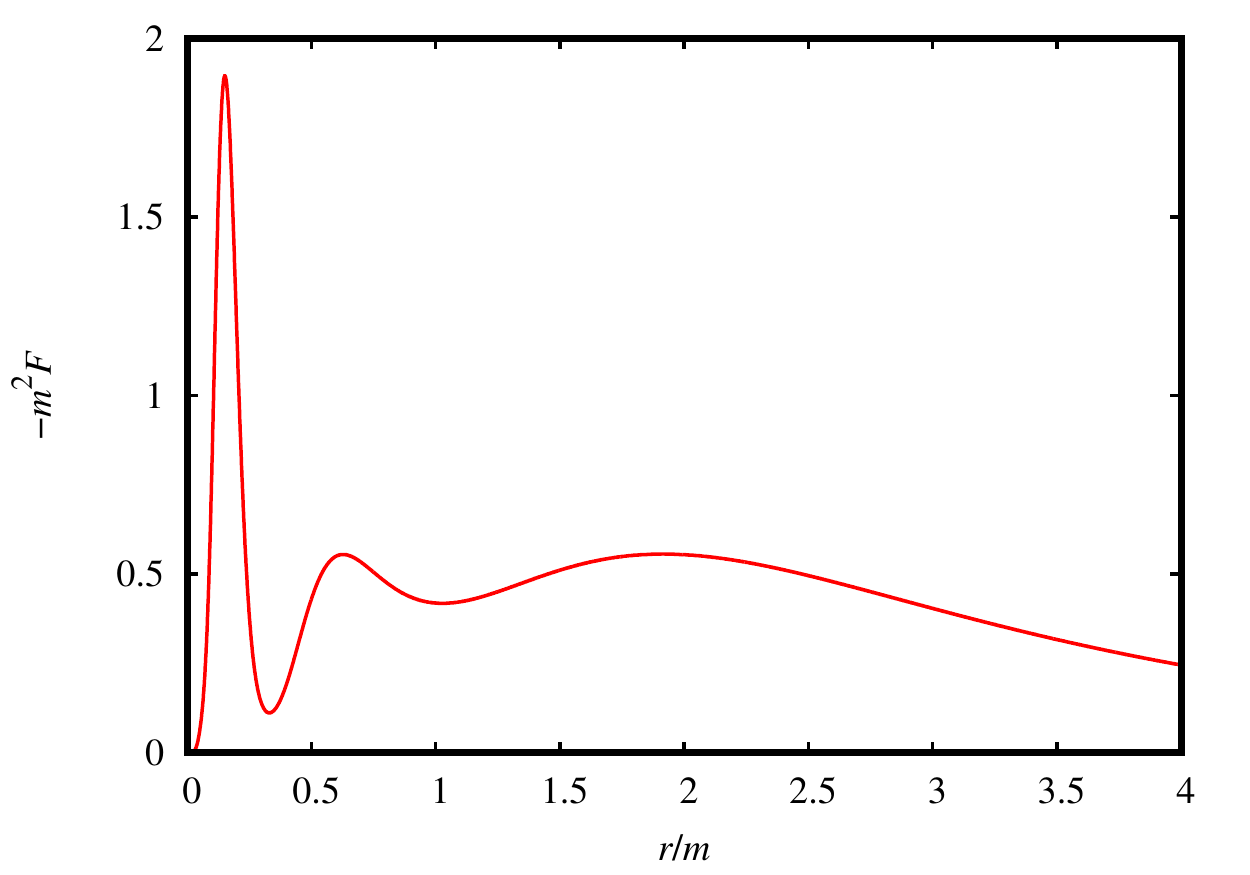}
	\caption{Graphical representation of $-F(r)$ to $q=1.07m$.}
	\label{Fr2}
\end{figure}

\begin{figure*}[htb!]
	\subfigure[]{\includegraphics[width=\columnwidth]{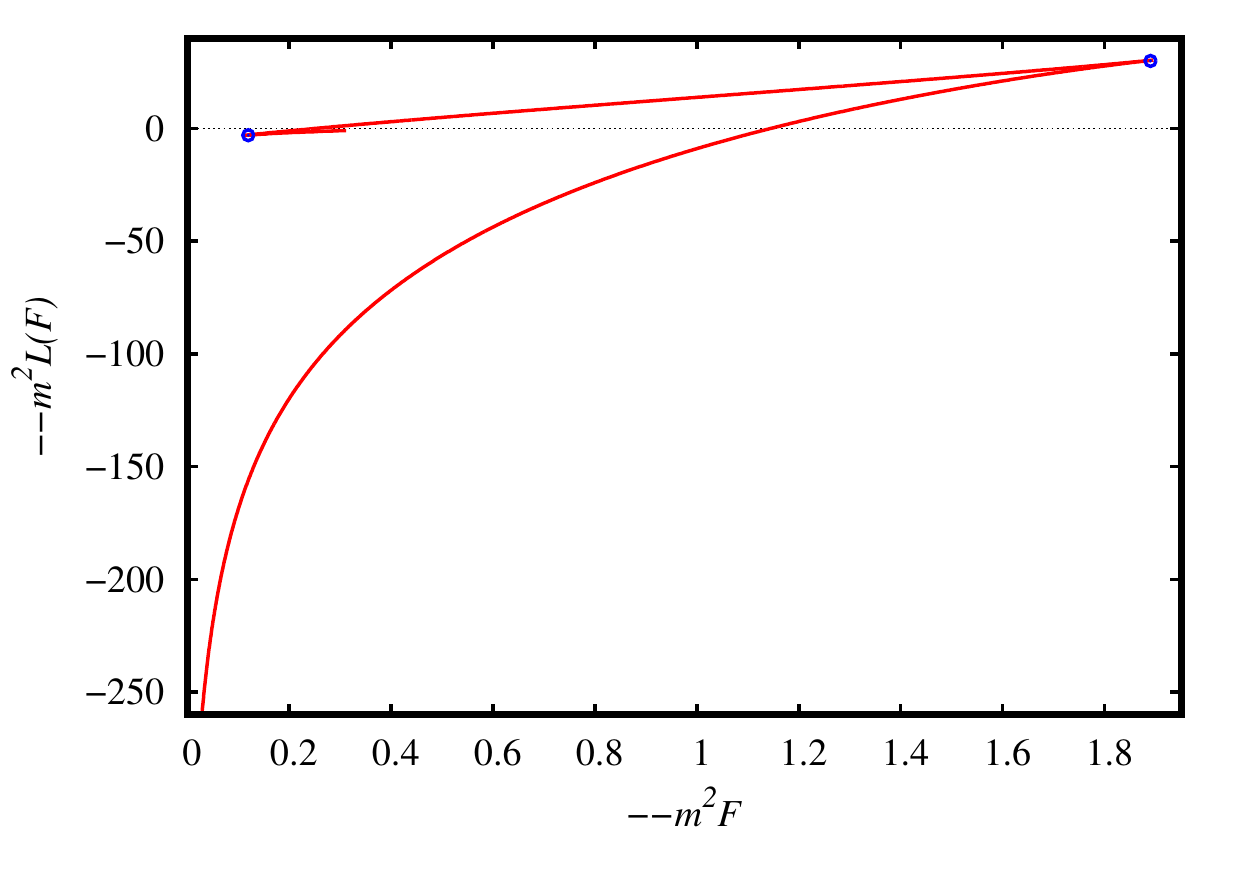}}
	\subfigure[]{\includegraphics[width=\columnwidth]{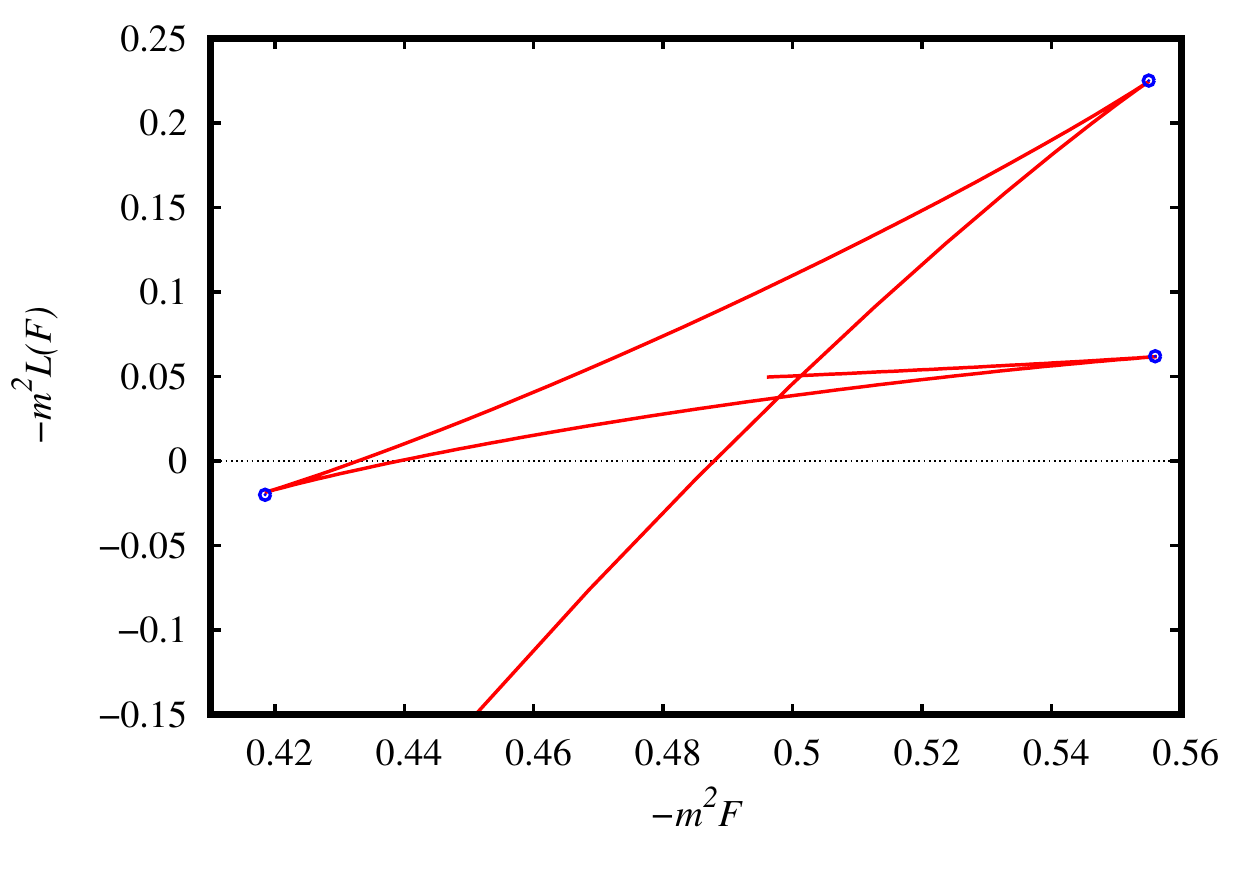}}
	\caption{Parametric representation of $-L(-F)$ to $q=1.07m$. The interval of $r$ is (a) $[0,0.45m]$ and (b) $(0.45m,2.5m]$.}
	\label{LxF4}
\end{figure*}

The asymptotic form of $L(r)$ and $F(r)$ are given by
\begin{eqnarray}
L(r\rightarrow \infty)&\sim &  -\frac{3 \left(4 m^2+q^2\right)}{r^4}+O(r^{-5}),\\
L(r\rightarrow 0)&\sim & \frac{16 \left(256+81 \pi ^4\right) m^4}{\pi ^4 q^6}-\frac{46656 m^5 r}{q^8},\nonumber\\ \\
F(r\rightarrow \infty)&\sim & -\frac{18 \left(4 m^2 +q^2\right)^2}{q^2 r^4}+O\left(r^{-5}\right),\\
F(r\rightarrow 0)&\sim & -\frac{120932352 m^{10} r^6}{q^{18}} .
\end{eqnarray}
The asymptotic dependence of $L(F)$ is 
\begin{eqnarray}
L(F)&\sim& F, r\rightarrow \infty,\\
L(F)&\sim&  \frac{16 \left(256+81 \pi ^4\right) m^4}{\pi ^4 q^6}\nonumber\\
&-&\frac{1296 \sqrt[3]{3m^{10}} \sqrt[6]{-2F} }{q^5},r\rightarrow 0.
\end{eqnarray}
We see that in the infinity of the radial coordinate, the electrodynamics behaves asymptotically as Maxwell but do not in the center of the solution. In Fig. \ref{Fr2}, we graphically represent the function $-F(r)$, where we may see five extremes, three maximums and two non null minimums. So we have five cusps in $-L(-F)$, which are represented in Fig. \ref{LxF4}.

\begin{figure}[htb!]
	\includegraphics[width=\columnwidth]{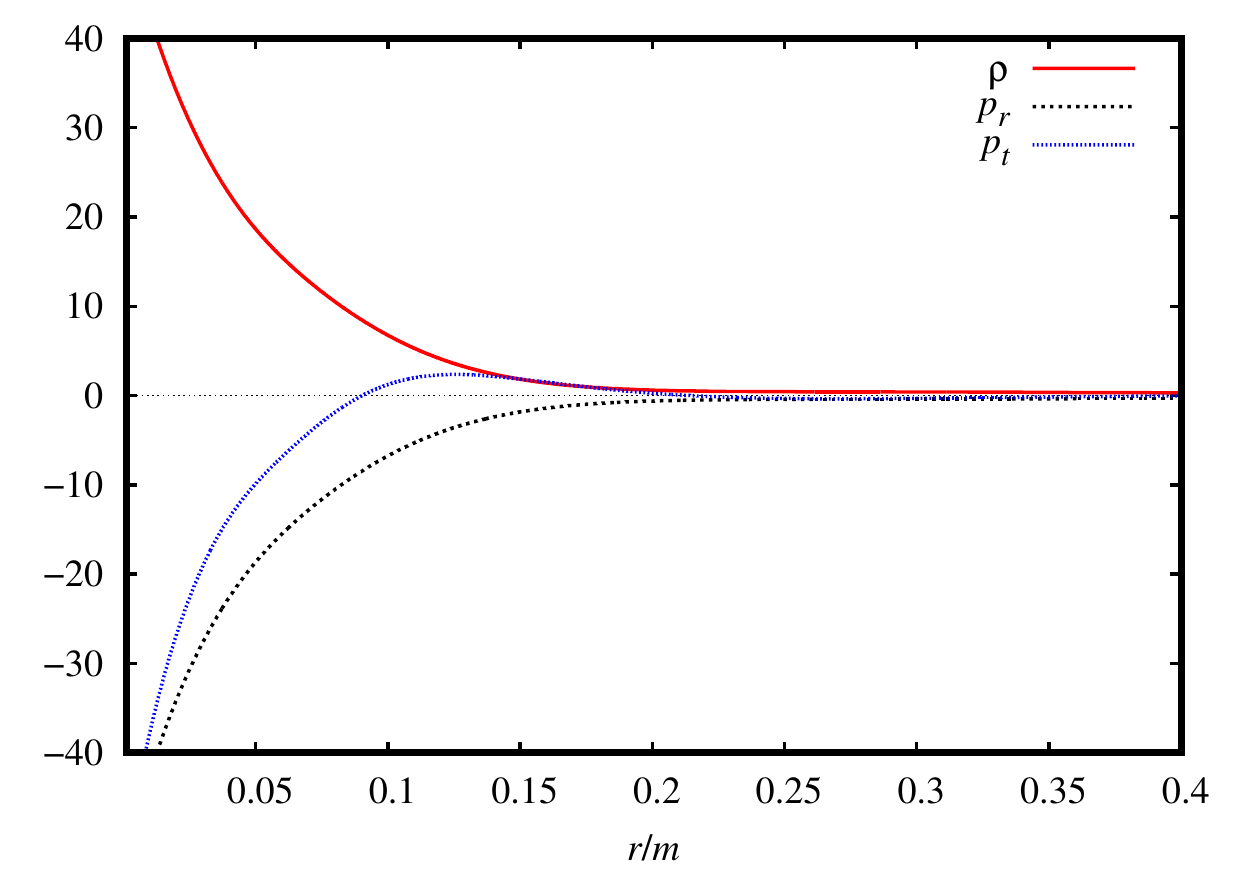}\vspace{-.1cm}
	\caption{Components of the stress-energy tensor to the solution \eqref{TS} for $q=m$.}
	\label{densSOL3}
\end{figure}

As this solution has a structure composed of several horizons, the components of the stress-energy tensor are analytically extensive. In Fig. \ref{densSOL3} we analyze the energy density, radial pressure and tangential pressure. As the examples before, we have the behavior of an anisotropic fluid, but, for regions close to $r=0$, is approximately isotropic.  

The asymptotic forms are
\begin{eqnarray}
\rho(r\rightarrow \infty)&\sim & \frac{3 \left(4 m^2+q^2\right)}{8 \pi  r^4}+O\left(r^{-5}\right)    ,\\
\rho(r\rightarrow 0)&\sim &\left(\frac{2 m^4 \left(256+81 \pi ^4\right)}{\pi ^5 q^6} +O\left(r\right)\right)+\nonumber\\
&+& e^{-\frac{q^2}{2 m r}} \left(\frac{q^2}{8 \pi  r^4}+O\left(r^{-2}\right)\right) ,\\
\omega_t(r\rightarrow \infty)&\sim & 1-\frac{144 m^2 q^2+96 m^4 +7 q^4}{36 m r \left(4 m^2+q^2\right)}+O(r^{-2}),\nonumber\\ \\
\omega_t(r\rightarrow 0)&\sim &-1. 
\end{eqnarray}

\begin{figure}[hbtp]
	\centering
	\includegraphics[width=\columnwidth]{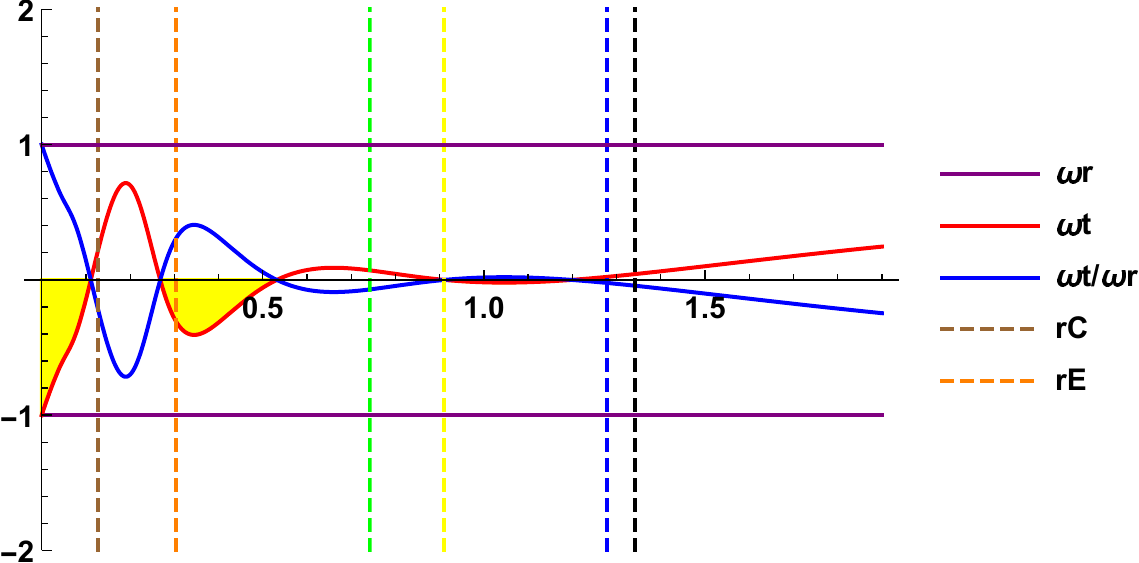}
	\caption{Graphical representation $\omega_{t}$, $\omega_{r}$, $\omega_{t}/\omega_{r}$ as a function of $r$ to $q=1.07m$. Each horizontal line represents a horizon.}
	\label{Omega_Balart-Culetu-Dymnikova_1}
\end{figure}

The energy density is not always positive, however, depending on the charge, we may impose the positivity.
When the energy density is positive $WEC_3$ is positive. The remains energy conditions should be evaluated through $\omega_{r,t}$. In Fig. \ref{Omega_Balart-Culetu-Dymnikova_1} we see that $\omega_{t}$ assumes negatives values, so that, also do $SEC_3$ and then the strong energy condition is always violated, as expected to regular solutions, while the other conditions are satisfied for some values of charge.

The method used here may be used to build solutions with even more horizons. For that, we need to only consider more terms in Eq. \eqref{fgeral} with different mass functions. Examples of mass functions may be found at \cite{Fan1,Balart:2014jia}, where, depending on the parameters chosen in the mass functions, we will have different solutions.

\subsection{Magnetically charged solution}

In this section we will cover a solution with up to four horizons magnetically charged. This type of solution does not have cusps in the graphical representation of the $L(F)$, as will be shown soon.

\begin{figure}[htb!]
	\includegraphics[width=\columnwidth]{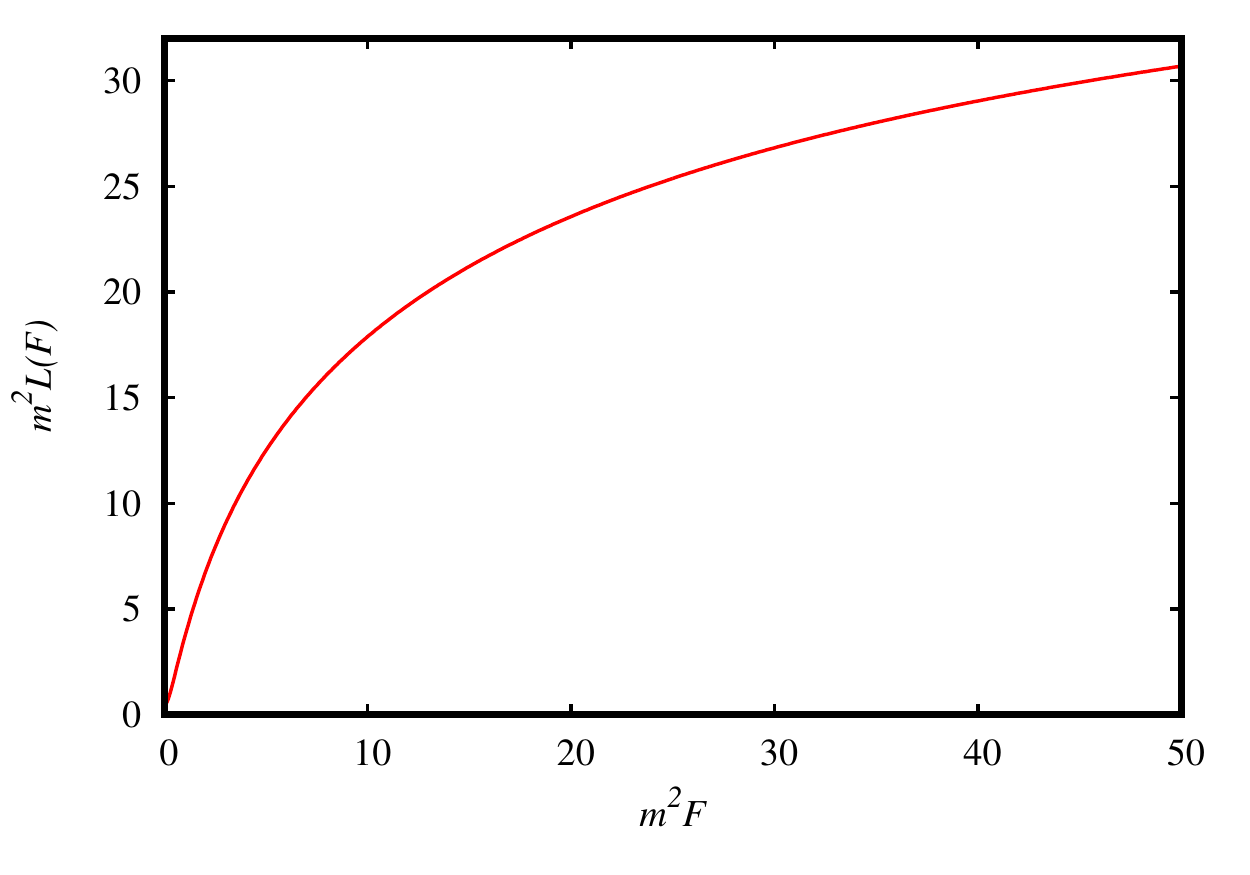}
	\caption{Behavior of $L(F)$ to $q_m=0.5m,l=m$ and $\beta=0.1m^2$.}
	\label{LxFmag}
\end{figure}

\begin{figure}[htb!]
	\includegraphics[width=\columnwidth]{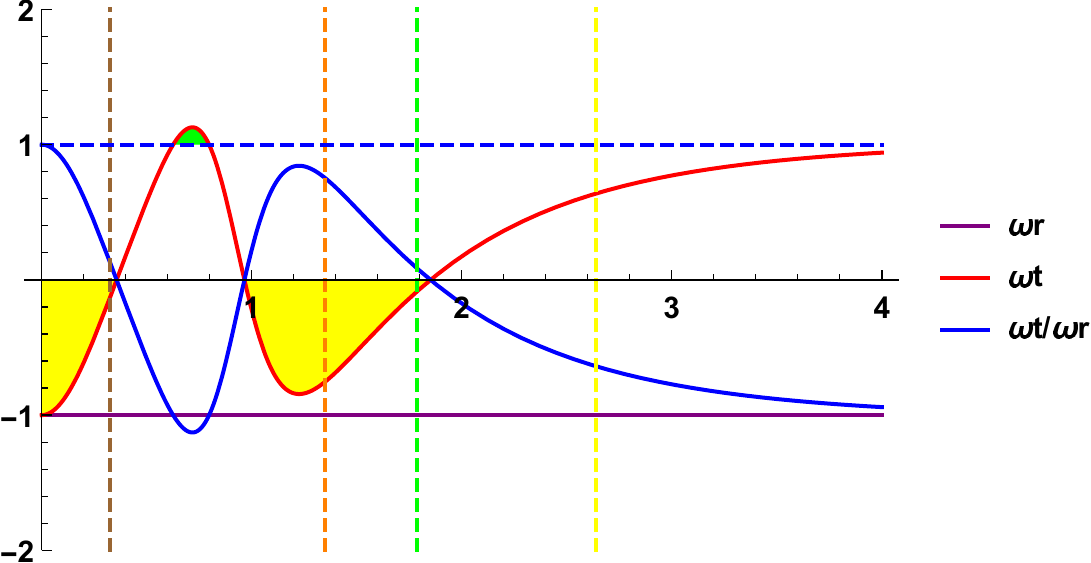}
	\caption{Graphical representation $\omega_{t}$, $\omega_{r}$, $\omega_{t}/\omega_{r}$ as a function of $r$ to $q=0.5m,l=m,\beta=0.1m^2$. Each horizontal line represents a horizon.}
	\label{omegaBK}
\end{figure}

Let's consider the Bardeen solution, $M_1(r)$, and the Kruglov solution, $M_5(r)$, that is given by \cite{Kruglov:2017xmb,Mazharimousavi:2019jja}
\begin{eqnarray}
M_5(r)&=&\frac{r^3h(r)}{r^3+2l^2h(r)}\,,\\
h(r)&=&\left[m+\frac{\left|q_m\right|^{3/2}}{2^{3/4}\sqrt[4]{\beta}}\tan^{-1}\left(\frac{2^{1/4}r}{\sqrt{\left|q_m\right|}\sqrt[4]{\beta}}\right)\right]\,,
\end{eqnarray}
where $q_m$ is the magnetic charge, $\beta$ is a real and positive parameter and $l$ is the fundamental length scale constant. The metric function is 
\begin{eqnarray}
f(r)=\left(1-\frac{2M_1(r)}{r}\right)\left(1-\frac{2M_5(r)}{r}\right)\,.
\end{eqnarray}
The asymptotic behavior is given by
\begin{eqnarray}
f(r\rightarrow \infty)&\sim & 1-\frac{4 m+\frac{\pi  \left|q_m\right|^{3/2}}{2^{3/4} \sqrt[4]{\beta }}}{r}+O(r^{-2}),\\
f(r\rightarrow 0)&\sim & 1-r^2 \left(\frac{1}{l^2}+\frac{2 m \sqrt{q_m^2}}{q_m^4}\right)+O\left(r^3\right).\nonumber\\
\end{eqnarray} 
The spacetime is asymptotically flat and has a de-Sitter core. We have four horizons in the interval of magnetic charge $0.342150825324m<q_m<4m/(3\sqrt{3})$ with $\beta=0.1m^2$ and $l=m$ (we will consider this values from now). The asymptotic behavior of the Kretschmann scalar is
\begin{eqnarray}
K(r\rightarrow 0)&\sim &\left(\frac{24}{l^4}+\frac{96 m \left|q_m\right|}{l^2 q_m^4}+\frac{96 m^2}{q_m^6}\right)\nonumber\\
&-&r^2 \Bigg(\frac{240 m \left|q_m\right|}{l^4 q_m^4}+\frac{480 m^2}{l^2 q_m^6}+\frac{360 m  \left|q_m\right|}{l^2 q_m^6}\nonumber\\
&+&\frac{720 m^2}{q_m^8}\Bigg)+O\left(r^3\right).
\end{eqnarray}
The asymptotic behavior of $L(F)$ is
\begin{eqnarray}
L(F\rightarrow \infty)&\sim &  \frac{3}{l^2}+\frac{6 m \left|q_m\right|}{q_m^4}\nonumber\\
&-&\frac{5 m \left(3 l^2+2 q_m^2\right)}{\sqrt{2} l^2 q_m^4 }F^{-1/2}+O(F^{-3/4}),\nonumber\\ \\
L(F\rightarrow 0)&\sim & \left(\frac{8 m^2}{q_m^2}+\frac{2 \sqrt[4]{2} \pi  m}{\sqrt[4]{\beta } \sqrt{\left|q_m\right|}}+2\right)F.
\end{eqnarray} 
In Fig. \ref{LxFmag}, we see that there is no cusps. The asymptotic behavior of the energy density and $\omega_t(r)$ are
\begin{eqnarray}
\rho(r\rightarrow \infty)&\sim & \frac{\frac{\sqrt[4]{2} \pi  m \left| q_m\right| ^{3/2}}{\sqrt[4]{\beta }}+q_m^2+4 m^2}{8 \pi  r^4}+O\left(r^{-5}\right)    ,\\
\rho(r\rightarrow 0)&\sim &\left(\frac{3}{8\pi l^2}+\frac{3 m \left|q_m\right|}{4\pi q_m^4}\right)\nonumber\\
&-&\frac{5 r^2 m \left|q_m\right| \left(3 l^2+2 q_m^2\right)}{8\pi l^2 q_m^6} +O\left(r^3\right),\\
\omega_t(r\rightarrow \infty)&\sim &  1+\frac{m q_m^2}{r \left(\sqrt[4]{2} \pi  \sqrt[4]{\frac{1}{\beta }} m \left| q_m\right| ^{3/2}+ q_m^2+4
	 m^2\right)}\nonumber\\&+&O\left(r^{-2}\right),\\
\omega_t(r\rightarrow 0)&\sim & -1+\frac{5 m r^2 \left(3 l^2+2 q_m^2\right)}{3 \left|q_m\right| \left(2 l^2 m \left|q_m\right|+q_m^4\right)}\nonumber\\&+&O\left(r^3\right). 
\end{eqnarray}
In Fig. \ref{omegaBK}, we represent the behavior of $\omega_{r,t}$ and we may see that $SEC_3$, yellow region, and $DEC_3$, green region, are violated. In particular, $DEC_3$ is violated to $r\rightarrow \infty$. 

\section{Conclusion}\label{sec5}

In this work we proposed a way to build solutions of regular black holes with multihorizons in general relativity. With that, we verified some properties of these solutions as regularity, energy conditions and electric field.

For the case of regular black holes with two horizons, we revised the Balart-Vagenas solution. This solution is regular, behaves asymptotically as Reissner-Nordstrom and has a de Sitter core with a constant Kretschmann scalar in the center and null in the infinity. WEC, NEC and DEC are satisfied in all points of the spacetime for any value of charge while SEC is violated inside the event horizon. The scalar $-F(-P)$ has non null a maximum at $P=512m^4/q^6$, and because of this, we have a cusp with two branches in the representation $L(F)$. Analytically, we obtained that the Lagrangian $L(F)$ behaves as Maxwell to $r\rightarrow \infty$ however do not to $r\rightarrow 0$. So that, as $F\rightarrow 0$ to $r\rightarrow 0$ and $r\rightarrow \infty$, the Lagrangian $L(F)$ has two different forms to the same $F$.

We formulate two examples with up to four horizons with electric charge. The first one, called the Bardeen-Culetu case, may present up to four horizons, if we have the value of the charge $ q <q ^ {BD} _ {ext} = [4m / (3 \sqrt {3})] $. If $ q = q ^ {BD} _ {ext} $ we have three horizons, and above that value up to $ q <q ^ {CL} _ {ext} = [2 m/ \sqrt{e}] $, we have two horizons. When $q=q^{CL}_{ext}$ we have only one horizon, and above that value, we have no more horizons. The solution is different from that of Bardeen and Culetu, with different energy density, different radial and tangential pressures, different electric field and nonlinear Lagrangian. The solution is regular with a constant Kretschmann scalar in the center. The scalar $-F(r)$ has nine extremes, 5 local maximums and 4 local null minimums. For which null minimum in $-F(r)$, the function $-L(-F)$ touch smoothly the axis $F=0$ and for which maximum we have a cusp in $-L(-F)$. Since the ratio between charges is $q^{CL}_{ext}/q^{BD}_{ext}=1.57581$, WEC and NEC may be satisfied for charges close to $q^{BD}_{ext}$, but DEC and SEC are always violated within the event horizon. The second example, called Balart-Culetu, has up to four horizons, just like the case of Bardeen-Culetu, depending only on the ratio between charge and mass. In this case, unlike the previous one, we may have a solution with four horizons that satisfy NEC, WEC, and DEC; the SEC is always violated within the event horizon. This is due to the ratio $q^{BL}_{ext}/q^{CL}_{ext}=1.09915$, which shows that the extreme charges are close. The solution behaves as Maxwell in the infinity and has a nonlinear behavior in the center. The scalar $-F(r)$ has two local maximums and a minimum. As $-F(r_{min})$ is not null, as the maximums, it represents a cusp in the Lagrangian $-L(-F)$, which has three cusps.
\par 
We also proposed a solution with up to six horizons, called Barlart-Culetu-Dymnikova. Likewise, this solution has the ratios between the extreme charges $q^{BL}_{ext}/q^{CL}_{ext}=1.09915$ and $q^{CL}_{ext}/q^{DC}_{ext}=1.13048 $, which shows how close they are. So we can have charges values close to $q^{DC}_{ext}=1.07304927103275m$, where NEC, WEC and DEC are satisfied, and SEC is violated within the event horizon. If we choose a solution as Bardeen, where the extreme charge is small in relation to the others, then the ratio between these charges would be appreciable, and therefore cannot satisfy the energy conditions. The scalar $-F(r)$ has three maximums and two non null minimum and, due to this fact, $-L(-F)$ has five cusps with six brunches.
\par 
We proposed a solution with up to four horizons but with magnetic charge, we called Bardeen-Kruglov solution. Different from the case with electric charge, $F$ just goes to zero in $r\rightarrow \infty$ where the Lagrangian does not behaves as Maxwell, as we already happen with the isolated Bardeen solution. As the scalar $F$ do not present the same value to different ranges of the radial coordinate, $-L(-F)$ presents no cusps. In this solution DEC is violated since $DEC_3$ presents negative values to $r\rightarrow \infty$ as the $SEC$ which is violated inside the event horizon once $SEC_3$ is negative in this region.
\par
From the results obtained, we can conclude the following: Given two or more solutions multiplied to form a general solution, where one of the solutions has an extreme charge with a very different value, or very close to the others, then it will not be possible that all energy conditions, except the SEC, be satisfied. Now, if the isolated solutions have an extreme charge reasonably close to another, it will be possible, for values close to the extreme, to satisfy all conditions except the SEC. In general, solutions can be built in such a way that the horizons depended only on the charge and mass ratio, it is also possible to write other physical quantities ($ L $, $ F $, $ L_ {F} $, ...) like this. Also $ q\rightarrow -q $ does not change the metric and other related quantities. It seems to us that all regular mass functions have a maximum of two horizons.
\par
This work opens up a new possibility of solutions for regular black holes with more than two horizons. We can have as perspectives the geodetic analysis, shadows, stability, causal structure and maximum space-time analytical extension of the solutions presented here. Also, to analyze Hawking temperature and thermodynamics of these black holes. The thermodynamic system of these solutions is much richer than the isolated solutions. We can also verify the phase transition, add the cosmological constant to an extended phase model, such as the van der Waals model.


\vspace{1cm}

{\bf Acknowledgements}: M. E. R.  thanks Conselho Nacional de Desenvolvimento Cient\'ifico e Tecnol\'ogico - CNPq, Brazil for partial financial support. This study was financed in part by the Coordenação de Aperfeiçoamento de Pessoal de Nível Superior - Brasil (CAPES) - Finance Code 001. The authors would like to thank the anonymous referee for his important suggestions that made it possible to improve the article.



\end{document}